\documentclass[10pt,aps,prc,floatfix,twocolumn,nofootinbib]{revtex4-1}

\usepackage{amsfonts}
\usepackage{amsmath}
\usepackage{amssymb}
\usepackage{bbold}
\usepackage{bm}
\usepackage{cellspace}
\usepackage{color}
\usepackage{enumerate}
\usepackage{graphicx}
\graphicspath{{Figures/}} 
\usepackage{hyperref} 
\usepackage{physics} 
\usepackage[figuresright]{rotating}
\usepackage{siunitx}
\usepackage[caption=false]{subfig} 

\newcommand{\chiEFT}{\ensuremath{\chi{\textrm{EFT}}}}
\newcommand{\LambdaBD}{{\Lambda_{\text{BD}}}}
\newcommand{\Trel}{\ensuremath{T_{\textrm{rel}}}}
\newcommand{\ataq}{a^{\dagger}_q a_q}
\newcommand{\abi}{\textit{ab initio}}

\newcommand{\mystrut}{\rule{0pt}{0.9\normalbaselineskip}}
\newcommand{\Klo}{K_{\text{lo}}}
\newcommand{\Khi}{K_{\text{hi}}}

\newcommand{\PP}{\ensuremath{P\mbox{--}P}}
\newcommand{\PQ}{\ensuremath{P\mbox{--}Q}}
\newcommand{\QP}{\ensuremath{Q\mbox{--}P}}
\newcommand{\QQ}{\ensuremath{Q\mbox{--}Q}}

\begin{document}

\title{Operator evolution from the similarity renormalization group \texorpdfstring{\\}{}
and the Magnus expansion}

\author{A.~J.~Tropiano$^{1}$, S.~K.~Bogner$^{2}$, R.~J.~Furnstahl$^{1}$}

\affiliation{%
$^1$\mbox{Department of Physics, The Ohio State University, Columbus, OH 43210, USA}  \\
$^2$\mbox{Facility for Rare Isotope Beams and Department of Physics and Astronomy,}  \\
\mbox{Michigan State University, East Lansing, MI 48824, USA}
}

\date{\today}

\begin{abstract}
The Magnus expansion is an efficient alternative to solving similarity renormalization group (SRG) flow equations with high-order, memory-intensive ordinary differential equation solvers. The numerical simplifications it offers for operator evolution are particularly valuable for in-medium SRG calculations, though challenges remain for difficult problems involving intruder states. Here we test the Magnus approach in an analogous but more accessible situation, which is the free-space SRG treatment of the spurious bound-states arising from a leading-order chiral effective field theory (EFT) potential with very high cutoffs. We show that the Magnus expansion passes these tests and then use the investigations as a springboard to address various aspects of operator evolution that have renewed relevance in the context of the scale and scheme dependence of nuclear processes. These aspects include SRG operator flow with band- versus block-diagonal generators, universality for chiral EFT Hamiltonians and associated operators with different regularization schemes, and the impact of factorization arising from scale separation. Implications for short-range correlations physics and the possibilities for reconciling high- and low-resolution treatments of nuclear structure and reactions are discussed.
\end{abstract}

\maketitle

\newpage

\section{Introduction}
\label{sec:intro}

Similarity renormalization group (SRG) transformations are a valuable tool for low-energy nuclear physics, whether applied in free space to soften input Hamiltonians for few- and many-body calculations, or for in-medium SRG (IMSRG) calculations that directly target the ground state or low-lying states in a given nucleus ~\cite{Bogner:2009bt,Furnstahl:2013oba,Hergert:2016iju}.
For both free-space and in-medium formulations, it is imperative that other operators are consistently and accurately evolved so that measurable quantities are left invariant.  
In the present work, we address the robustness of the Magnus expansion as a method to solve free-space SRG flow equations, and examine other issues of SRG operator evolution in light of the proliferation of new chiral EFT (\chiEFT) interactions~\cite{Epelbaum:2014efa,Gezerlis:2014zia,Piarulli:2014bda,Ekstrom:2015rta,Carlsson:2015vda,Reinert:2017usi,Ekstrom:2017koy,Entem:2017gor}, the scale dependence of short-range-correlation (SRC) physics~\cite{Hen:2016kwk,Weiss:2016obx,Cruz-Torres:2019fum,Schmidt:2020kcl}, recent interest in high-cutoff effective field theories (EFTs) and renormalization~\cite{Tews:2018sbi,Hammer:2019poc,vanKolck:2020llt,Tews:2020hgp}, and the universality of evolved operators~\cite{Dainton:2013axa,Arriola:2016fkr}.

The SRG decouples low- and high-momentum scales in a Hamiltonian by applying a continuous unitary transformation $U(s)$, where $s=0 \rightarrow \infty$ is the flow parameter~\cite{Bogner:2006pc}.
An evolved operator is given by
\begin{eqnarray}
	\label{eq:srg_operator}
	O(s) = U(s) O(0) U^{\dagger}(s),
\end{eqnarray}
where $O(0)$ is the initial operator.
Because $U(s)$ is unitary, matrix elements of the operator in evolved states are preserved.
An evolved operator can be found by solving a differential flow equation obtained by taking the derivative of Eq.~\eqref{eq:srg_operator},
\begin{eqnarray}
	\label{eq:srg_flow}
	\frac{dO(s)}{ds} = [\eta(s), O(s)],
\end{eqnarray}
where $\eta(s)=\frac{dU(s)}{ds} U^{\dagger}(s) = -\eta^{\dagger}(s)$ is 
the anti-hermitian SRG generator.
For the free-space SRG, the generator is typically defined as a commutator, $\eta(s) = [G, H(s)]$, where $G$ specifies the type of flow.
The choice of $G$ determines the pattern of decoupling in the Hamiltonian.

By setting $G=H_D(s)$, the diagonal of the Hamiltonian, the Hamiltonian is driven to band-diagonal form~\cite{Wegner:1994ab}.
In low-energy nuclear physics, $G$ is usually taken to be the relative kinetic energy, \Trel; i.e., the diagonal of the potential is not included in $G$. 
In most nuclear physics applications these two choices give the same evolved operators. 
But in exceptional cases involving evolution across bound states, which we consider in the next section, the two band-diagonal choices can have drastically different behaviors~\cite{Glazek:2008pg,Wendt:2011qj}. 
For band-diagonal decoupling, it is convenient to define $\lambda \equiv s^{-1/4}$, which roughly measures the width of the band-diagonal in the decoupled Hamiltonian~\cite{Anderson:2010aq}.

For block-diagonal decoupling~\cite{Anderson:2008mu,Szpigel:2016rbt}, $G$ is formed by splitting the Hamiltonian into low- and high-momentum sub-blocks as specified by a momentum separation scale $\LambdaBD$,
\begin{eqnarray}
	\label{eq:block_diag_g}
	G =
	\begin{pmatrix}
		PH(s)P & 0 \\
		0 & QH(s)Q \\
	\end{pmatrix}
	\equiv H_{BD}(s).
\end{eqnarray}
Here $P$ and $Q$ are low- and high-momentum projection operators.
In momentum space, the projection operators are step functions defined by the sharp cutoff $\LambdaBD$, although smoothed versions are also possible and may be preferred in some applications to avoid numerical artifacts~\cite{Anderson:2008mu}. 
These transformations are similar to V\textsubscript{low $k$} transformations~\cite{Bogner:2001gq,Bogner:2001jn,Bogner:2003wn} but keep the high-momentum matrix elements non-zero, maintaining a unitary transformation in the full space.
Complete decoupling of the blocks is in principle only reached in the $s \rightarrow \infty$ limit. 
In practice it is sufficient to solve the flow equation \eqref{eq:srg_flow} up to some finite value of $s$ with a high-order ODE solver such that the remaining ``neck'' between blocks is much narrower than $\LambdaBD$.

The SRG procedure can be implemented by solving the flow equation Eq.~\eqref{eq:srg_flow} for the evolved Hamiltonian simultaneously with other operators of interest.
However, one can also solve Eq.~\eqref{eq:srg_flow} exclusively for the evolved Hamiltonian and build the unitary transformation directly using the eigenvectors of the evolved and initial Hamiltonians (as is done in Sec.~\ref{sec:evolution_other_operators}).
Another approach is to solve the following equation for the unitary transformation
\begin{eqnarray}
	\label{eq:unitary_trans}
	\frac{dU(s)}{ds} = \eta(s) U(s),
\end{eqnarray}
which arises in an intermediate step in deriving Eq.~\eqref{eq:srg_flow}.
This is the starting point in the Magnus expansion implementation of the SRG.

The Magnus expansion gives us the capability to solve for the SRG unitary transformation with negligible violations of unitarity from numerically solving the ODEs,%
\footnote{There is a small numerical violation of unitarity in the standard approach to solving SRG equations due to accumulated time-step errors. With the Magnus expansion, unitarity is preserved to much higher precision because of the form of the transformation, as detailed in Sec.~\ref{sec:magnus_expansion_formalism}.}
after which it can be applied to any other operator of interest~\cite{Morris:2015yna}.
By utilizing an exponential parameterization for the transformation, $U(s)=e^{\Omega(s)}$, Eq.~\eqref{eq:unitary_trans} is recast as a flow equation for the anti-Hermitian operator $\Omega(s)$.
The solution of the flow equation for $\Omega(s)$ permits the use of cheap low-order ODE methods since the exponentiated operator is still unitary even if it has accumulated non-negligible time-step errors~\cite{Morris:2015yna}.
The Magnus expansion also offers important advantages over the direct solution of Eq.~\eqref{eq:unitary_trans} in Fock space, where practical calculations require operators to be truncated at the $a-$body level ($a<A$).
For instance, even if the Magnus flow equations are truncated at the two-body level, the resulting unitary transformation contains higher-body components from the exponentiation of $\Omega$.%
\footnote{This is similar to the advantages of truncated Coupled Cluster theory calculations relative to truncated Configuration Interaction calculations.}

Due to these advantages, most large-scale IMSRG calculations now utilize the Magnus expansion.
There are still open problems though.
For instance, in applications of the IMSRG to derive effective valence shell model Hamiltonians in multi-shell valence spaces, intruder states, which are low-lying states whose wave functions are dominated by high-energy configurations outside the model space, can severely distort low-energy properties or even prevent the flow from converging.
It is not yet fully understood how the IMSRG procedure evolves intruder state systems, though it appears that induced three- and higher-body operators rapidly grow in size for such systems, destroying the cluster hierarchy ($2N \gg 3N \gg 4N \gg \ldots$) in the evolved Hamiltonian~\cite{Stroberg:2019mxo}.

Interestingly, there is an analog to the intruder state problem in the much simpler two-nucleon problem.
In spin-triplet channels and at leading-order (LO) in \chiEFT, taking the EFT cutoff to high values can result in spurious, deep-bound states due to the highly singular short-ranged tensor force from one-pion exchange. 
In principle, these deep-bound states are not a problem because they are outside the range of the EFT.
In practice, there are subtleties analogous to the intruder state problem when one attempts to soften such Hamiltonians with free space SRG evolution.
In Ref.~\cite{Wendt:2011qj}, it was shown that band-diagonal SRG decoupling of NN potentials in partial waves with spurious bound states fails for the standard $G=\Trel$ generator, as the flow forces the deep bound state into the low-momentum sector.
As a result, there is no decoupling of high- and low-momentum physics, and the evolved interactions become increasingly singular at low momentum.
In contrast, the Wegner generator $G=H_{D}$ succeeds at depositing the spurious state(s) along the diagonal in the high-momentum sector, which is more natural as it allows a clean decoupling of high- and low-momentum physics. 
Since these findings were for the direct solution of Eq.~\eqref{eq:srg_flow}, this  provides a good test case for the Magnus approach and we document its performance in detail.
More generally, there has been renewed interest in studying chiral interactions at high cutoffs~\cite{Tews:2018sbi}. 
These high-cutoff chiral potentials provide us a laboratory to explore the effects of the SRG generator on decoupling, universality, and SRCs.
These issues also inform the behavior of standard \chiEFT\ potentials.

While interactions from \chiEFT\ have become the standard choice for \abi\ calculations of nuclei, they are not unique, even when restricted to the commonly used Weinberg power counting, because of many choices for regularization schemes and fitting protocols and even degrees of freedom (i.e., with or without Deltas).
In the early applications of \chiEFT\ potentials to nuclei, these choices were not explored but in recent years there has been a proliferation of nucleon-nucleon (NN) potentials and associated three-nucleon forces (e.g., see Refs.~\cite{Epelbaum:2014efa,Gezerlis:2014zia,Piarulli:2014bda,Ekstrom:2015rta,Carlsson:2015vda,Reinert:2017usi,Ekstrom:2017koy,Entem:2017gor}).
This diversification motivates us to revisit SRG operator evolution.
Past studies were limited to phenomenological interactions or a single class of chiral interactions (namely the non-local-regulated potentials from Refs.~\cite{Entem:2003ft} or \cite{Epelbaum:2004fk}).
Here we examine the fate of scheme dependence for new-generation NN potentials and associated operators as they are evolved to lower resolutions.

One intriguing aspect is universality.
By virtue of fitting to the same data or phase shifts, different \chiEFT\ potentials generate close to the same $S$-matrix in the energy range where there is a good fit; that is, the potentials are phase equivalent in that range.
However, matrix elements of the potentials in momentum space differ significantly based on the EFT order and the choice of regulator function and cutoff (scale and scheme dependence). 
Nevertheless, it has been observed that SRG transformations drive different NN potentials toward the same low-momentum matrix elements; in particular, this flow to universality is seen up to the momentum value of phase inequivalence~\cite{Bogner:2003wn,Bogner:2001gq,Dainton:2013axa}. 
We examine whether universality holds for modern chiral potentials but also address universality for other operators evolving under the corresponding SRG transformations.
This has implications for the analysis of reactions at different resolution scales~\cite{Furnstahl:2013dsa,More:2015tpa,More:2017syr}.

The question of whether non-Hamiltonian operators decouple or take universal forms has not been fully addressed in the literature.%
\footnote{Note that if the wave functions are decoupled, it is not necessary for the operators themselves to decouple to get decoupled matrix elements. See examples below.}
In fact, the decoupling of matrix elements does not necessarily result for other operators as it does for the Hamiltonian.
In previous work \cite{Anderson:2010aq,Bogner:2012zm}, it was found that SRG evolution induces low-momentum contributions in high-momentum operators and changes low-momentum operators very little, as might be expected from general EFT considerations.
We investigate whether this is a general trend of the SRG for a wider selection of potentials and SRG generators, explicitly analyze the nature of the evolution for representative high-momentum and low-momentum operators, and relate these observations to the high-resolution picture of SRCs and the role of factorization.

The plan of the paper is as follows.
We first revisit the high-cutoff problem, and test the Magnus approach in Sec.~\ref{sec:high_cutoffs_magnus_exp}.
We consider evolution of new-generation NN Hamiltonians in Sec.~\ref{sec:srg_evolution_nn_potentials} and then turn to other operators in Sec.~\ref{sec:evolution_other_operators}.
Our conclusions and outlook are summarized in Sec.~\ref{sec:summary}.

\section{High cutoffs and the Magnus expansion}
\label{sec:high_cutoffs_magnus_exp}

\subsection{High cutoffs and spurious bound states}
\label{sec:high_cutoffs}

\begin{figure*}[tbh]
    \includegraphics[clip,width=0.8\textwidth]{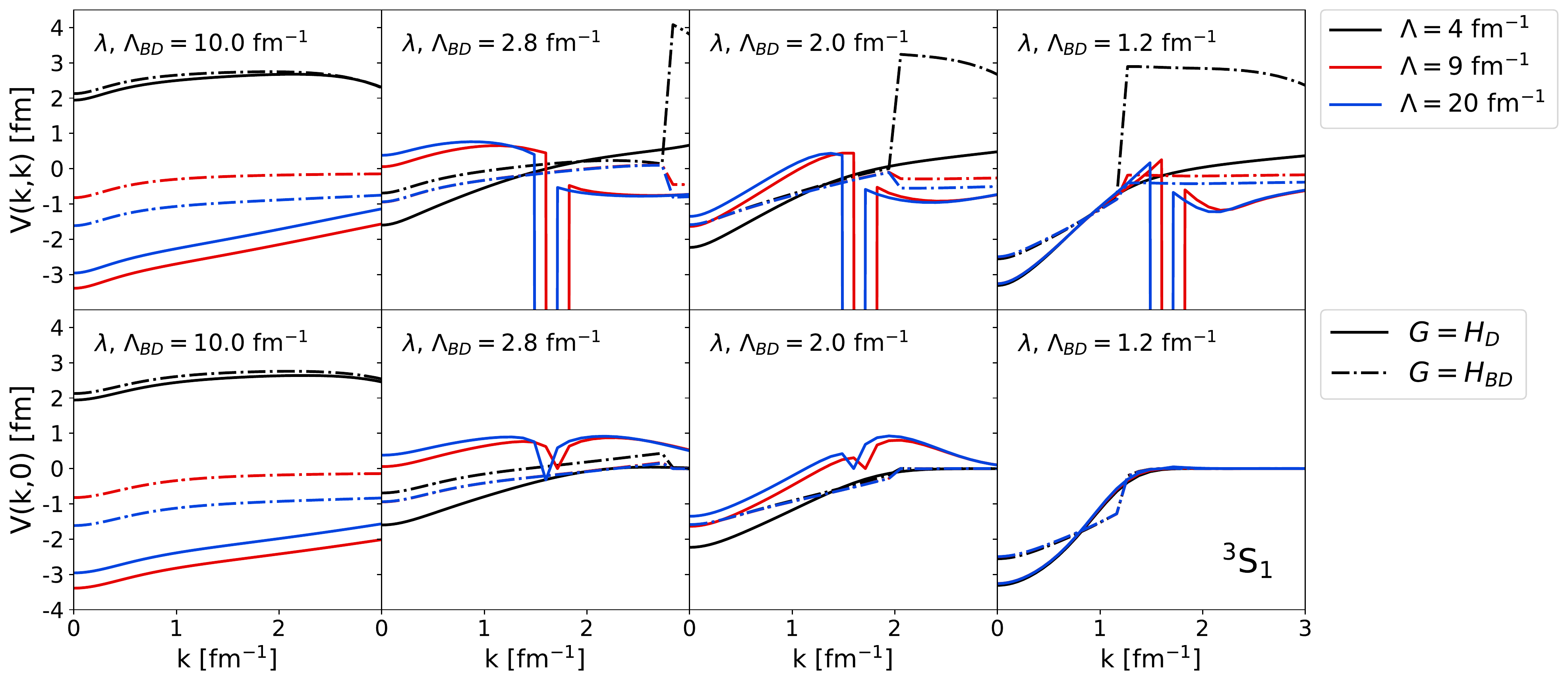}
	\caption{Diagonal and far off-diagonal matrix elements of non-local LO potentials at cutoffs $\Lambda=4$ (black), $9$ (red) and $20$\,fm$^{-1}$ (blue), SRG-evolving left to right under transformations with Wegner (solid) and block-diagonal (dashed) generators in the $^3$S$_1$ channel.
	We vary the SRG flow parameter $\lambda$ for Wegner evolution and fix it at $\lambda=1.2$\,fm$^{-1}$ for block-diagonal evolution. The decoupling scale in the block-diagonal generator is denoted by $\LambdaBD$.}
	\label{fig:potential_slices_high_cutoffs}
\end{figure*}

The \chiEFT\ potentials used in most \abi\ nuclear calculations are not renormalizable in the sense that dependence on the regulator is not suppressed by taking the momentum cutoff increasingly high (or low, if a coordinate-space regulator).
However, Nogga \textit{et al.}~\cite{Nogga:2005hy} showed that the LO version of these interactions, with promoted counterterms in some channels, \emph{is} renormalizable in this sense.
There is active work on renormalizable power counting for \chiEFT\ beyond LO (see references cited in \cite{Tews:2018sbi}).

The LO theory at high cutoff is a useful laboratory for testing the SRG (as well as providing insight into the evolution of SRC physics, see Sec.~\ref{sec:evolution_other_operators}).
It features the appearance of spurious, deeply bound states in some channels, which is ultraviolet physics beyond the range of the EFT, and thus does not violate EFT principles.
However, these present a major challenge to the SRG.
Wendt \textit{et al.} studied SRG band-diagonal transformations of high-cutoff LO potentials~\cite{Wendt:2011qj} and showed that channels with spurious deep-bound states did not automatically exhibit the expected decoupling and universality of the potential if the conventional SRG generator is used.
In particular, the spurious bound state(s) is driven from high to low momentum in the evolved potential when applying transformations with $G=\Trel$.
The observables remain unchanged because the transformation is still unitary, but the potential and wave functions are altered significantly by the presence of the spurious bound state at low momentum. 
In contrast, if the Wegner generator is used, the spurious state(s) is decoupled, subsequently yielding universality in low-momentum matrix elements of the potential.

In Fig.~\ref{fig:potential_slices_high_cutoffs} we show SRG band- and block-diagonal evolution of high-cutoff non-local potentials at LO, 
which consists of one-pion exchange and a contact interaction.
We restrict our attention to the $^3$S$_1$--$^3$S$_1$ sub-block of the coupled $^3$S$_1$--$^3$D$_1$ channel (note that spurious, deeply bound states only appear in spin-triplet channels~\cite{Nogga:2005hy}).
The contact interaction is determined by fitting the associated low-energy constant to $E_{\text{lab}}=10$ MeV phase shift data.
In the diagonal matrix elements of Fig.~\ref{fig:potential_slices_high_cutoffs}, we see a steep drop-off in the Wegner transformed potentials around $k \approx 1.6-1.75$\,fm$^{-1}$.
This corresponds to the decoupled spurious bound state ($\varepsilon \approx -2000$ MeV).
The value of momentum where the spurious bound state decouples is a scheme-dependent quantity that is sensitive to how momentum space is discretized (the momentum mesh).
Due to this dependence, we have been unable to predict the value of $k$ at which the spurious state decouples.
However, when the spurious state is decoupled outside the low-momentum part of the potential, the Wegner evolution collapses the low-momentum matrix elements of the different potentials to the same mesh-independent values in accordance with universality.
There is no drop-off in the $\Lambda=4$\,fm$^{-1}$ potential as it has no spurious state.

We also see universality for the block-diagonal evolved matrix elements as before, but there is no noticeable influence from the spurious bound state.
This is due to the band-diagonal generator locally decoupling the matrix elements whereas the block-diagonal generator cleanly separates the potential into a low-momentum sub-block and a high-momentum sub-block.
In the limit $\lambda \rightarrow 0$ with $\LambdaBD$ sufficiently low, the spurious deep-bound state(s) is contained entirely in the high-momentum sub-block.
We verified this by diagonalizing the sub-blocks separately to see which one contained the spurious bound state.
We identify $\LambdaBD \approx 4.5$\,fm$^{-1}$ as the approximate value at which the spurious bound state switches from the low-momentum sub-block to the high-momentum sub-block.
Thus, the block-diagonal transformations decouple the spurious state(s) at a higher value of momentum than the Wegner transformations, which isolates the physical states more effectively.
Tests with different meshes found the same value of momentum, suggesting a scheme-independent result.
However, we do not have an analytic understanding of these scales.

We can draw a loose analogy to intruder states corrupting low-energy physics in IMSRG calculations with spurious bound states corrupting universality in SRG-evolved potentials.
From the analysis so far, it is evident that the choice of SRG generator is important in properly decoupling the high-momentum spurious state from low-momentum physics.
It would be interesting to generalize this conclusion to an $A$-body system and analyze how different generators deal with intruder states.
However, we must first verify that the Magnus approach is the same as the conventional SRG approach for difficult systems because of its use in IMSRG calculations.
In the following sub-sections, we present the Magnus approach and compare to the conventional SRG using high-cutoff potentials with spurious states as a test case.

\subsection{The Magnus expansion: Formalism}
\label{sec:magnus_expansion_formalism}

We briefly review the formalism of the Magnus expansion and its use in the SRG.
Mathematically speaking, the Magnus expansion is a method for solving an initial value problem associated with a linear ordinary differential equation.
Formal details of the Magnus expansion are discussed in \cite{Blanes:2009ab}.
We will introduce the Magnus expansion in the context of SRG operator evolution.

\begin{figure*}[tbh]
	\includegraphics[clip,width=0.8\textwidth]{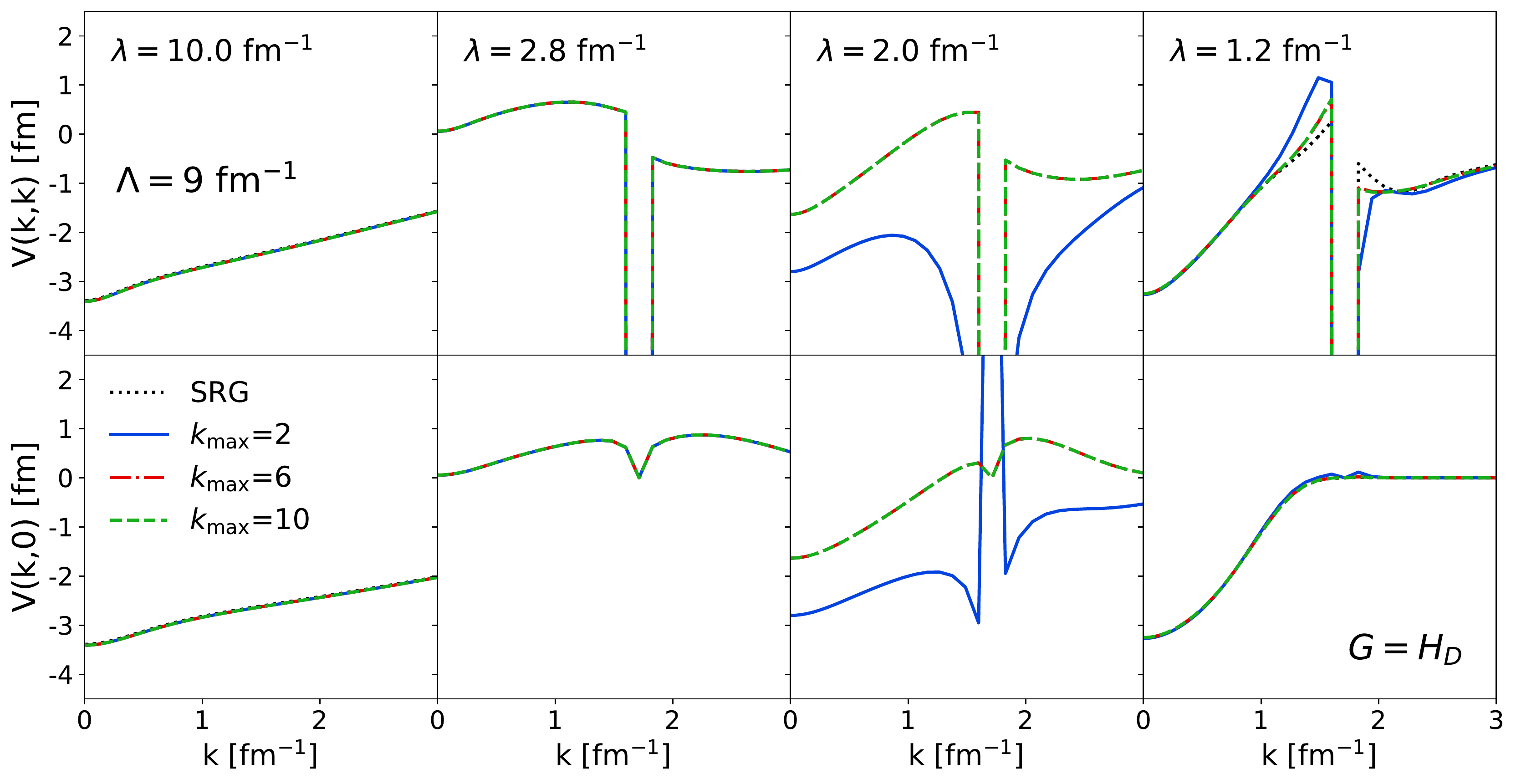}%
	\caption{Comparison of SRG- and Magnus-evolved diagonal and far off-diagonal matrix elements of the non-local LO potential (9\,fm$^{-1}$) in the $^3$S$_1$ channel for several truncations $k_{\rm max}$ in the Magnus sum \eqref{eq:magnus_omega}. Here we evolve in $\lambda$ left to right using the Wegner generator $G=H_D$.}
	\label{fig:potential_slices_high_cutoffs_Wegner}
\end{figure*}
\begin{figure*}[tbh]
	\includegraphics[clip,width=0.8\textwidth]{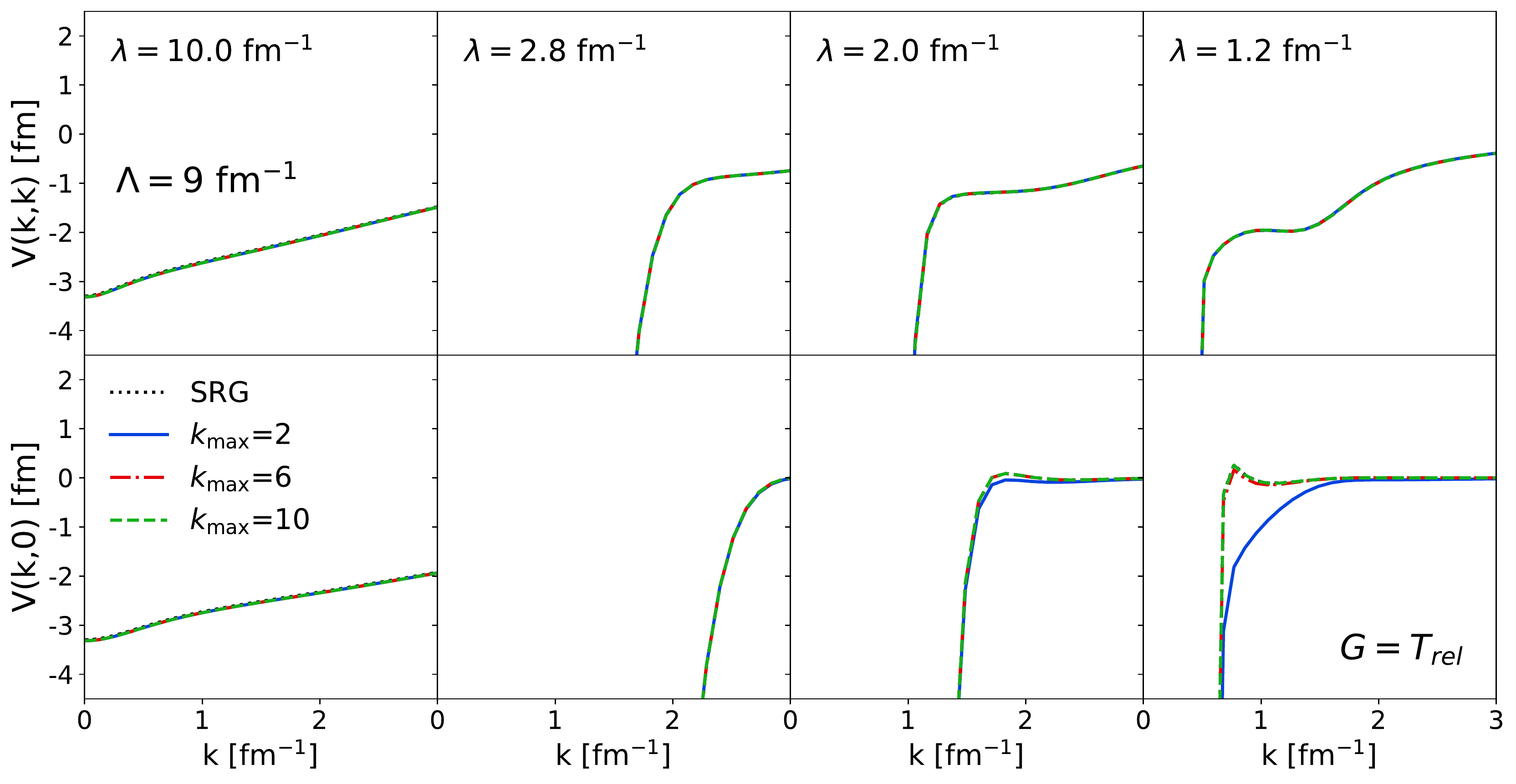}%
	\caption{Same as in Fig.~\ref{fig:potential_slices_high_cutoffs_Wegner} but with $G=\Trel$.}
	\label{fig:potential_slices_high_cutoffs_T}
\end{figure*}

We can solve Eq.~\eqref{eq:unitary_trans} with a solution $U(s)=e^{\Omega(s)}$ where $\Omega^{\dagger}(s) = -\Omega(s)$, and $\Omega(0) = 0$.
$\Omega(s)$ is expanded as a power series in $\eta(s)$:
\begin{eqnarray}
    \label{eq:magnus_expansion}
    \Omega(s) = \sum_{n=1}^{\infty} \Omega_n(s),
\end{eqnarray}
where the terms of the series are given by integral expressions involving $\eta(s)$,
\begin{align}
    \label{eq:omega_terms}
    \Omega_1(s) &= \int_0^s\, ds_1 \eta(s_1), \nonumber \\        
    \Omega_2(s) &= \frac{1}{2} \int_0^s ds_1 \int_0^{s_1} ds_2 \, [\eta(s_1), \eta(s_2)], \\
    & \quad\vdots \nonumber
\end{align}
Equation~\eqref{eq:magnus_expansion} is referred to as the Magnus expansion.
(Again, see \cite{Blanes:2009ab, Magnus:1954zz} for further
details.)
We avoid computing the integral terms $\Omega_n(s)$ since it requires storing $\eta(s)$ over a range of $s$ values, which is impractical for large-scale calculations.
We focus instead on the derivative of $\Omega(s)$,
\begin{eqnarray}
	\label{eq:magnus_omega}
	\frac{d\Omega(s)}{ds} = \sum_{k=0}^{\infty} \frac{B_k}{k!} ad_{\Omega}^{k}(\eta),
\end{eqnarray}
where $B_k$ are the Bernoulli numbers, $ad_{\Omega}^{0}(\eta)=\eta(s)$, and $ad_{\Omega}^{k}(\eta)=[\Omega(s),ad_{\Omega}^{k-1}(\eta)]$.
We integrate this differential equation to find $\Omega(s)$ and evaluate the unitary transformation.
Then the evolved operator can be evaluated with the Baker-Cambell-Hausdorff formula~\cite{Morris:2015yna},
\begin{eqnarray}
	\label{eq:bch}
	O(s) = e^{\Omega(s)} O(0) e^{-\Omega(s)} = \sum_{k=0}^{\infty} \frac{1}{k!} ad_{\Omega}^{k}(O).
\end{eqnarray}
As $k \rightarrow \infty$ in both sums in Eqs.~\eqref{eq:magnus_omega} and \eqref{eq:bch}, the Magnus transformation matches the SRG transformation exactly.%
\footnote{Note that this equivalence is exact only if both series converge and the ODEs in Eq.~\eqref{eq:magnus_omega} are solved exactly.}
We investigate several truncations $k_{\rm max}$ in Eq.~\eqref{eq:magnus_omega} and take many terms, $k_{\rm max} \sim 25$, in Eq.~\eqref{eq:bch}.

\begin{figure*}[tbh]
	\includegraphics[clip,width=0.67\textwidth]{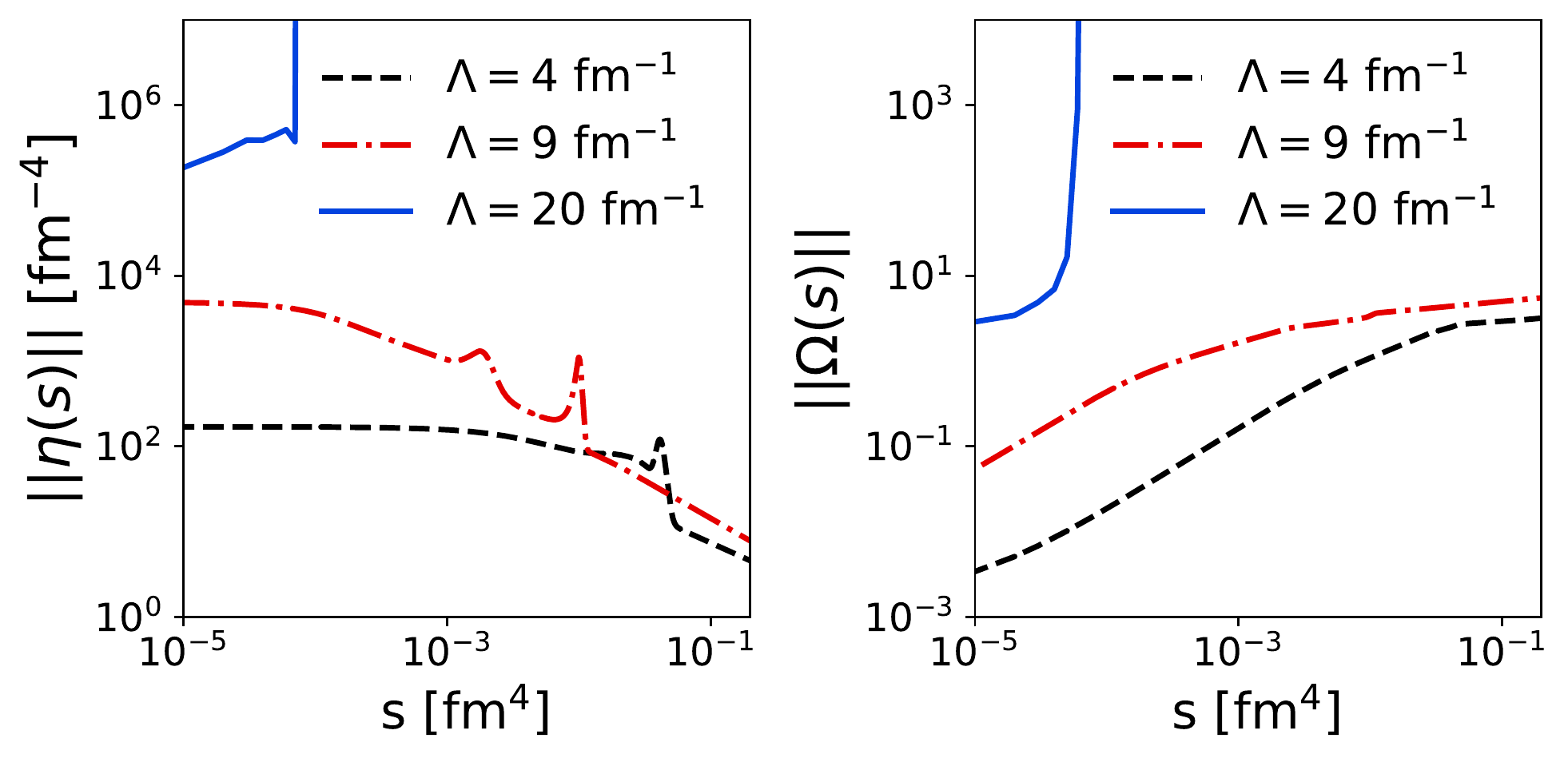}%
	\caption{Frobenius norms of $\eta(s)$ and $\Omega(s)$ from Magnus-evolving the high-cutoff LO potentials in the $^3$S$_1$--$^3$D$_1$ coupled channel: $\Lambda=4$ (black dashed), $9$ (red dash-dotted) and $20$\,fm$^{-1}$ (blue solid) where $G=H_D$ and $k_{\rm max}=6$.}
	\label{fig:eta_omega_norms}
\end{figure*}

There are significant advantages to the Magnus implementation in IMSRG calculations.
In the conventional approach, the numerical error associated with solving Eq.~\eqref{eq:srg_flow} accumulates directly in the operator and can distort the eigenvalues of the transformed Hamiltonian.
To guard against this, one must use a high-order ODE solver, which can become prohibitive for large-scale calculations due to the memory-intensive nature of such solvers.  
In the Magnus implementation, unitarity is guaranteed by the form of $U(s)$.
One can solve Eq.~\eqref{eq:magnus_omega} with a low-order stepping method with a substantially lower memory footprint, which nevertheless preserves the eigenvalues exactly while still decoupling as desired.
Here we demonstrate this advantage by applying the Magnus implementation using the first-order Euler step-method.
Note that the execution time for Magnus applied to free-space SRG is roughly the same as for conventional SRG evolution, because the more complicated evaluations in Eq.~\eqref{eq:magnus_omega} are offset by the fewer evaluations needed with a low-order ODE solver.

The second major advantage involves the evolution of multiple operators.
In many situations, one may be interested in evolving several operators at a time.
In the standard procedure, we would have another set of coupled equations in Eq.~\eqref{eq:srg_flow}, drastically increasing memory usage.
Each additional operator increases the set of equations - say $N$ equations - by another factor of $N$.
In the Magnus approach, one only needs $\Omega(s)$ to consistently evolve several operators via explicit construction of $U(s)=e^{\Omega(s)}$.
While operator evolution is not an issue for NN evolution, the capability to calculate $U(s)$ directly is crucial in IMSRG calculations where the model space can be very large.
While it is possible to solve for $U(s)$ by directly integrating Eq.~\eqref{eq:unitary_trans}, this suffers from the same memory limitations as Eq.~\eqref{eq:srg_flow} due to the necessity of a high-order ODE solver to guard against the loss of unitarity.

\subsection{The Magnus expansion: Results}
\label{sec:magnus_expansion_results}

We compare the SRG evolution of the non-local LO potential with cutoff at 9\,fm$^{-1}$ using the conventional approach and the Magnus approach.
At this particular cutoff, the potential has one spurious bound state in the $^3$S$_1$--$^3$D$_1$ coupled channel of about $-2000$ MeV in addition to the deuteron bound-state energy. 
Figures~\ref{fig:potential_slices_high_cutoffs_Wegner} and \ref{fig:potential_slices_high_cutoffs_T} show the diagonal and far off-diagonal matrix elements of the evolving potential using both methods at several different truncations $k_{\rm max}$ for Magnus-evolution for the two band-diagonal generators, $G=H_D$ and \Trel, respectively.
In both cases, as we take higher values of $k_{\rm max}$ the Magnus evolution approaches the SRG despite the presence of a spurious bound state.
The agreement is rather poor for the lowest truncation shown, $k_{\rm max}=2$.
Although the observables for the Magnus-evolved potentials are still unaltered independent of $k_{\rm max}$, the presence of the decoupled spurious bound state has effects on the flow to band-diagonal form.
That is, there is more variation with respect to $k_{\rm max}$ in band-diagonal decoupling of these potential matrix elements.

We have tested other, softer potentials such as the lower cutoff of 4\,fm$^{-1}$ and higher-order chiral potentials, and found that the Magnus implementation always works as intended.
The Magnus implementation nearly matches the SRG results in all cases where small differences come from the difference in ODE solver and truncations in the Magnus approach.
Thus, we only show results for the high cutoff of 9\,fm$^{-1}$.

In some cases, $\eta(s)$ grows as $s$ increases, leading to convergence issues in the Magnus expansion.
When $\eta(s)$ begins increasing, $\Omega(s)$ grows prohibitively large.
In Ref.~\cite{Blanes:2009ab} the convergence of the Magnus expansion is described in terms of the Frobenius norm of $\eta(s)$, stating that convergence is satisfied if $\int_0^S ||\eta(s)|| ds < r_c$ over an interval $0 < s < S$, where $r_c = \pi$ for calculations involving real matrices.

In Fig.~\ref{fig:eta_omega_norms} we show the Frobenius norms of $\eta(s)$ and $\Omega(s)$ for the three high-cutoff potentials tested using $G=H_D$.
The convergence issue arises for $\Lambda=20$\,fm$^{-1}$ at $s \sim 10^{-4}$ where $||\eta(s)||$ and subsequently $||\Omega(s)||$ jump several orders of magnitude.
However, the problem is completely avoided when the block-diagonal generator is used.
We have tested different Magnus truncations $k_{\rm max}$ and Euler method step-sizes and found the same behavior.

Overall, the Magnus implementation reproduces the generator-dependent SRG behavior for high-cutoff potentials, where the universality of the different potentials is achieved with the Wegner generator but not the relative kinetic energy generator.
We note that the block-diagonal generator decouples the spurious bound state(s) at a much higher momentum value than the Wegner band-diagonal generator and still flows to a universal form in the low-momentum matrix elements.
Although we fixed $k_{\rm max}$ in our results, one could use an adaptive method of selecting $k_{\rm max}$ values at each step in $s$ where criteria is based on flow to band- or block-diagonal form.
One could also truncate Eq.~\eqref{eq:magnus_omega} when the Frobenius matrix norm of the $k^{\rm th}$ term is significantly smaller than the matrix norm of the $0^{\rm th}$ term (see Ref.~\cite{Morris:2015yna} for further details).

From the convergence standpoint, the initial interaction and generator $\eta(s)$ clearly play a significant role in how the Magnus implementation works.
This should be no surprise from how $\eta(s)$ is defined in terms of the Hamiltonian.
In connection to the Magnus expansion in the IMSRG context, similar convergence issues arise for intruder state problems~\cite{Stroberg:2019mxo}.
At least for the NN system, our results imply that the choice of SRG generator can play a significant role in overcoming the issues stemming from intruder states, though we leave this as work for a future study.

\section{SRG evolution of NN potentials}
\label{sec:srg_evolution_nn_potentials}

\subsection{Modern chiral NN potentials}
\label{subsec:modern_chiral}

Next we extend our analysis of the SRG evolution of high-cutoff LO potentials to include higher-order chiral potentials.
In Ref.~\cite{Dainton:2013axa}, the conditions under which different potentials are driven to universal low-momentum matrix elements were studied.
Here we apply both band- and block-diagonal transformations to several newer chiral potentials, also focusing on universality.
For band-diagonal evolution, we use the Wegner generator, $G=H_D$, instead of \Trel, which was used in \cite{Dainton:2013axa}.
For these interactions with cutoffs of order 2--3\,fm$^{-1}$, the two band-diagonal choices are essentially equivalent, unlike for the potentials considered in the previous section when the cutoff was above 4\,fm$^{-1}$.
We restrict our analysis in this section and the following section to the typical SRG approach as results with the Magnus implementation and standard SRG are indistinguishable for soft potentials.

We will consider three representative potentials: the N$^4$LO potential with 500\,MeV cutoff from Ref.~\cite{Entem:2017gor} (denoted EMN N$^4$LO), the N$^4$LO potential with 450\,MeV cutoff from Ref.~\cite{Reinert:2017usi} (denoted RKE N$^4$LO), and the N$^2$LO potential with 1\,fm cutoff from Ref.~\cite{Gezerlis:2014zia} (denoted Gezerlis N$^2$LO).
These three potentials differ in the regulator functions applied to the contact and pion-exchange terms. 

\begin{figure*}[htb]
	\includegraphics[clip,width=0.85\textwidth]{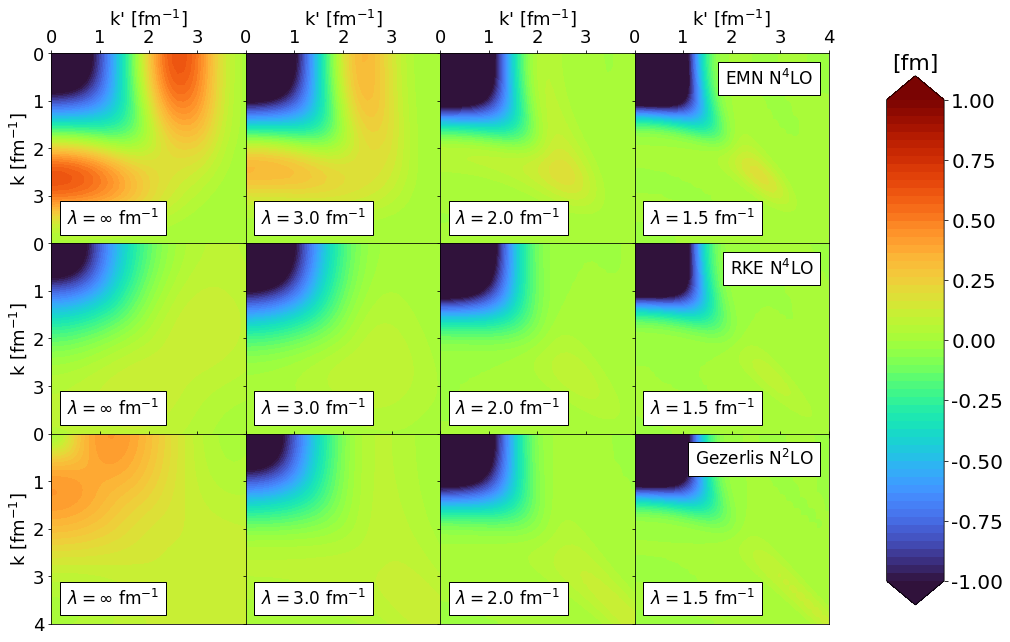}%
	\caption{Momentum-space matrix elements of the EMN N$^4$LO 500 MeV, RKE N$^4$LO 450 MeV, and Gezerlis \textit{et al.}~N$^2$LO 1 fm potentials SRG-evolved in $\lambda$ with the Wegner generator in the coupled $^3$S$_1$--$^3$D$_1$ channel (only $^3$S$_1$ is shown here.)}
	\label{fig:potential_contours_3S1_Wegner}
\end{figure*}
\begin{figure*}[tbh]
    \includegraphics[clip,width=0.85\textwidth]{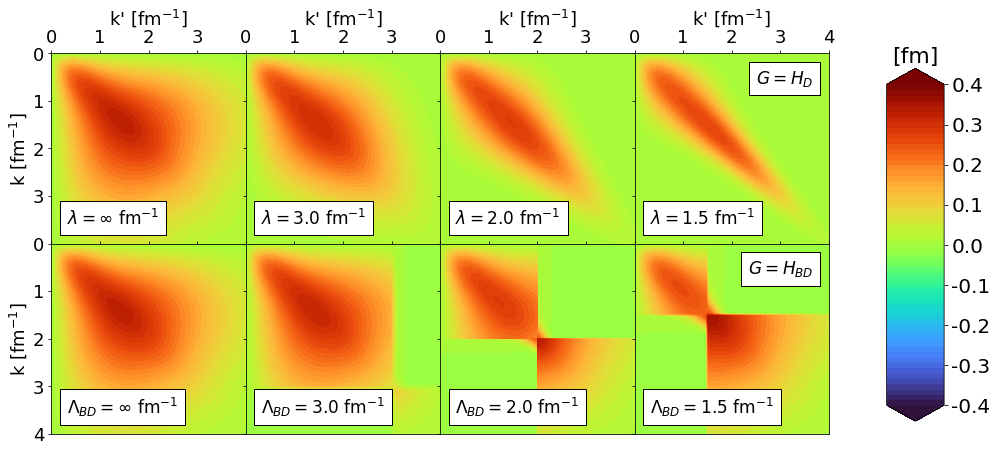}%
	\caption{Matrix elements of the RKE N$^4$LO 450 MeV potential SRG-evolving left to right under transformations with Wegner and block-diagonal generators in the $^1$P$_1$ channel. We vary the SRG flow parameter $\lambda$ for Wegner evolution and fix it at $\lambda=1$\,fm$^{-1}$ for block-diagonal evolution. The decoupling scale in the block-diagonal generator is denoted by $\LambdaBD$.}
	\label{fig:potential_contours_1P1_RKE}
\end{figure*}
\begin{figure*}[tbh]
	\subfloat[]{%
	\raisebox{.5in}{%
	\includegraphics[clip,width=0.25\textwidth]{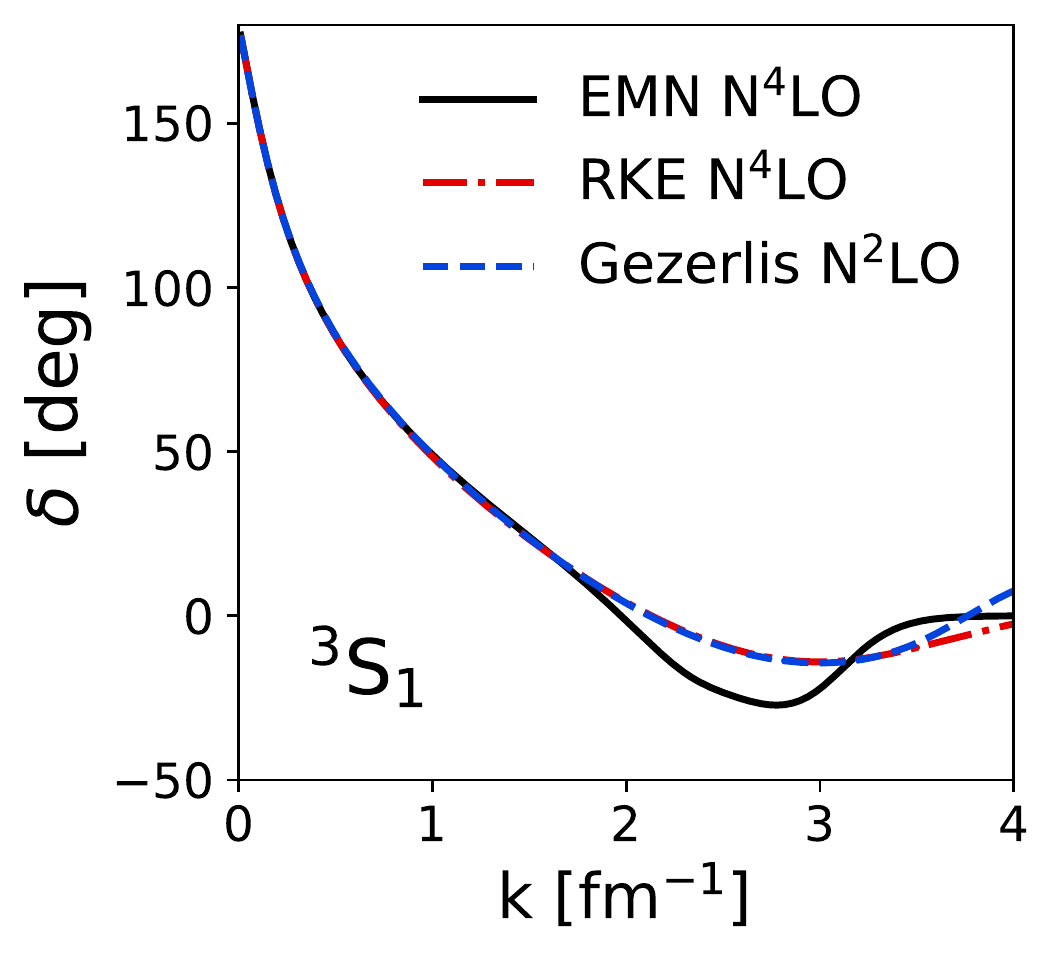}}}
	\hfill
	\subfloat[]{%
	\includegraphics[clip,width=0.72\textwidth]{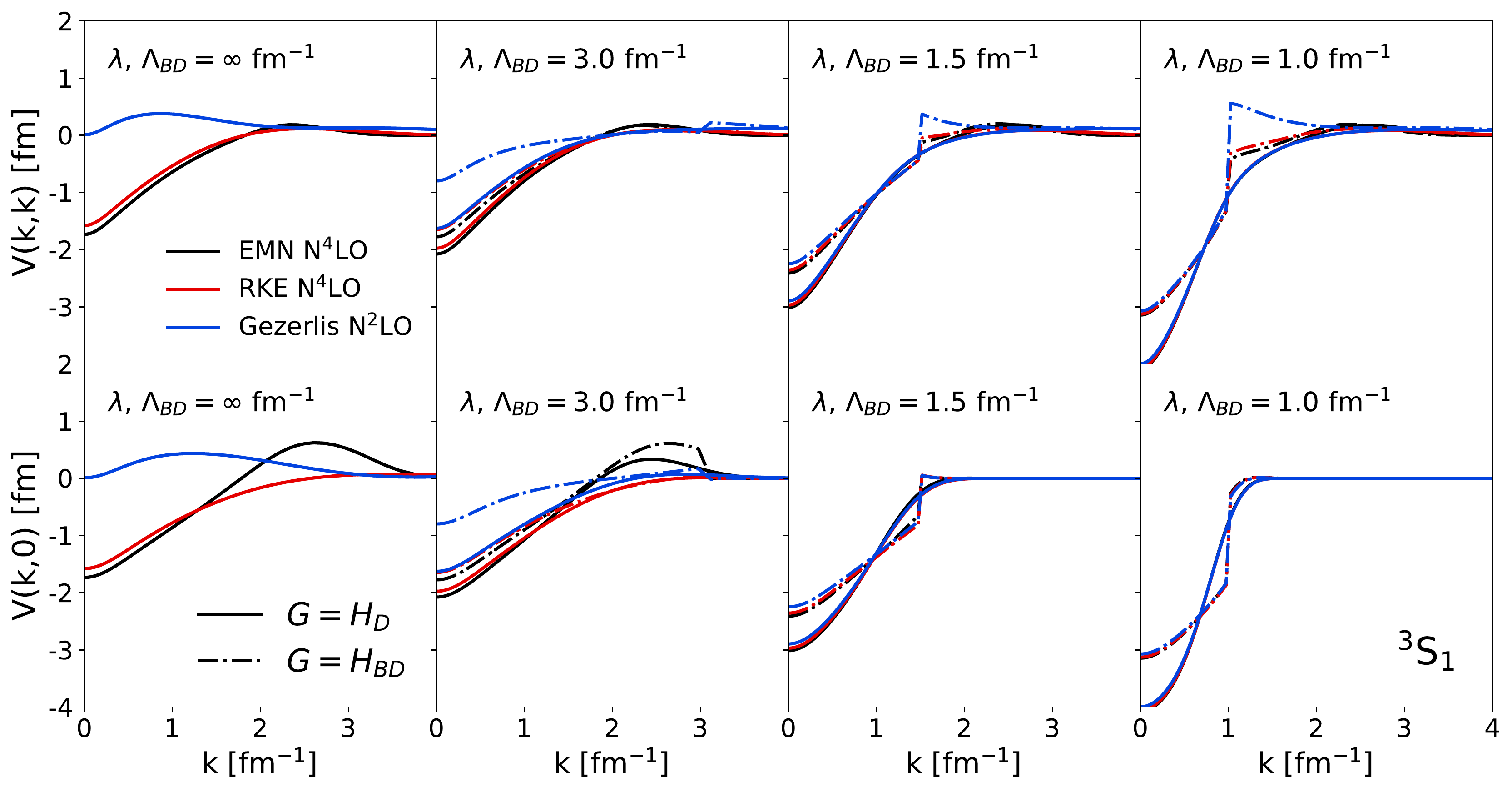}%
	}
	\caption{(a) $^3$S$_1$ phase shifts for the EMN N$^4$LO 500 MeV (solid black), RKE N$^4$LO 450 MeV (red dash-dotted), and Gezerlis \textit{et al.}~N$^2$LO 1 fm (blue dashed) potentials. (b) Diagonal and far off-diagonal matrix elements of the EMN N$^4$LO 500 MeV (black), RKE N$^4$LO 450 MeV (red) and Gezerlis \textit{et al.}~N$^2$LO 1 fm (blue) potentials SRG-evolved with Wegner (solid) and block-diagonal (dashed) generators in the $^3$S$_1$ channel. For block-diagonal evolution, we fix $\lambda=1$\,fm$^{-1}$.}
	\label{fig:potential_universality_3S1}
\end{figure*}

The EMN N$^4$LO interaction is a non-local potential where both contact and pion-exchange interactions feature a non-local regulator function of the form exp$[-(k/\Lambda)^{2n}-(k'/\Lambda)^{2n}]$, where $\Lambda$ is the momentum-space cutoff and $n$ is an integer. 
A non-local regulator function for pion-exchange contributions can introduce regulator artifacts by distorting the known analytic structure of the NN scattering amplitudes near threshold for cutoffs $\Lambda$ lower than the breakdown scale $\Lambda_b$~\cite{Epelbaum:2014efa}.
Semi-local chiral potentials have been introduced to reduce regulator artifacts, such as the RKE potentials.
Here, a local regulator function is applied for the long-range interactions in momentum space, while a non-local regulator function is used for the short-range interactions.
Non-local interactions are generally not suitable for continuum quantum Monte Carlo methods, motivating the need for fully local chiral potentials.
The Gezerlis \textit{et al.}~N$^2$LO potential is an example of a local interaction where both the long-range and short-range terms have a local regulator function in coordinate space.

These chiral interactions give the same low-energy phase shifts but the matrix elements of the potential are often completely different.
We show band-diagonal SRG evolution of the three potentials in the $^3$S$_1$ channel in Fig.~\ref{fig:potential_contours_3S1_Wegner}.
On the left-hand column where $\lambda=6$\,fm$^{-1}$, the three potentials differ dramatically.
Further along the SRG evolution (right-hand side), the potentials are driven to band-diagonal form where the upper left corner of the contours, corresponding to low-momentum matrix elements, become close to the same.

Figure~\ref{fig:potential_contours_1P1_RKE} shows the SRG-evolved RKE N$^4$LO (450 MeV cutoff) potential in the $^1$P$_1$ partial wave channel for band- and block-diagonal SRG generators on the top and bottom rows, respectively.
We continue to evolve to band-diagonal form with respect to the parameter $\lambda$, but for the block-diagonal generator, we label sub-plots with the parameter $\LambdaBD$ that characterizes the sharp cutoff in decoupling the low- and high-momentum matrix elements.
For complete block-diagonal decoupling, one should take $s \rightarrow \infty$, which corresponds to $\lambda \rightarrow 0$.
This is difficult to carry out in practice since the ODEs become stiff, so we stop the evolution at $\lambda=1$\,fm$^{-1}$.
We see a small non-zero width in between the sub-blocks due to the non-zero value of $\lambda$, but when the width is this small the sub-blocks are effectively decoupled.
With block-diagonal decoupling, one can truncate the Hamiltonian at the chosen value of $\LambdaBD$, separately diagonalize each sub-block, and retain all eigenvalues to high accuracy.
We have tested representative cases and found the same eigenvalues to better than $0.1$\% for both the low- and high-momentum sub-blocks.

\begin{figure*}[tbh]
    \includegraphics[clip,clip,width=0.75\textwidth]{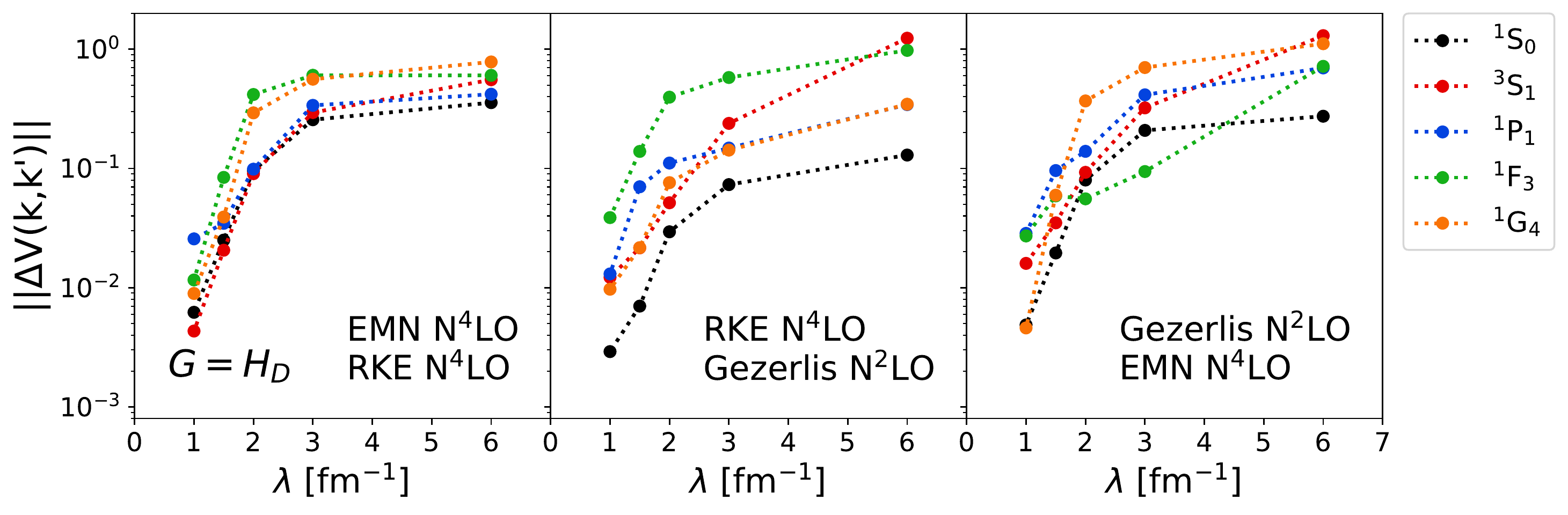}%
    \caption{Frobenius norm of the difference of two SRG-evolved potentials for several partial wave channels, $^1$S$_0$ (black), $^3$S$_1$ (red), $^1$P$_1$ (blue), $^1$F$_3$ (green), and $^1$G$_4$ (orange), comparing the three default potentials, EMN N$^4$LO 500 MeV, RKE N$^4$LO 450 MeV, and Gezerlis \textit{et al.}~N$^2$LO 1 fm.}
    \label{fig:universality_test_with_norm_Wegner}
\end{figure*}
\begin{figure*}[tbh]
    \includegraphics[clip,clip,width=0.75\textwidth]{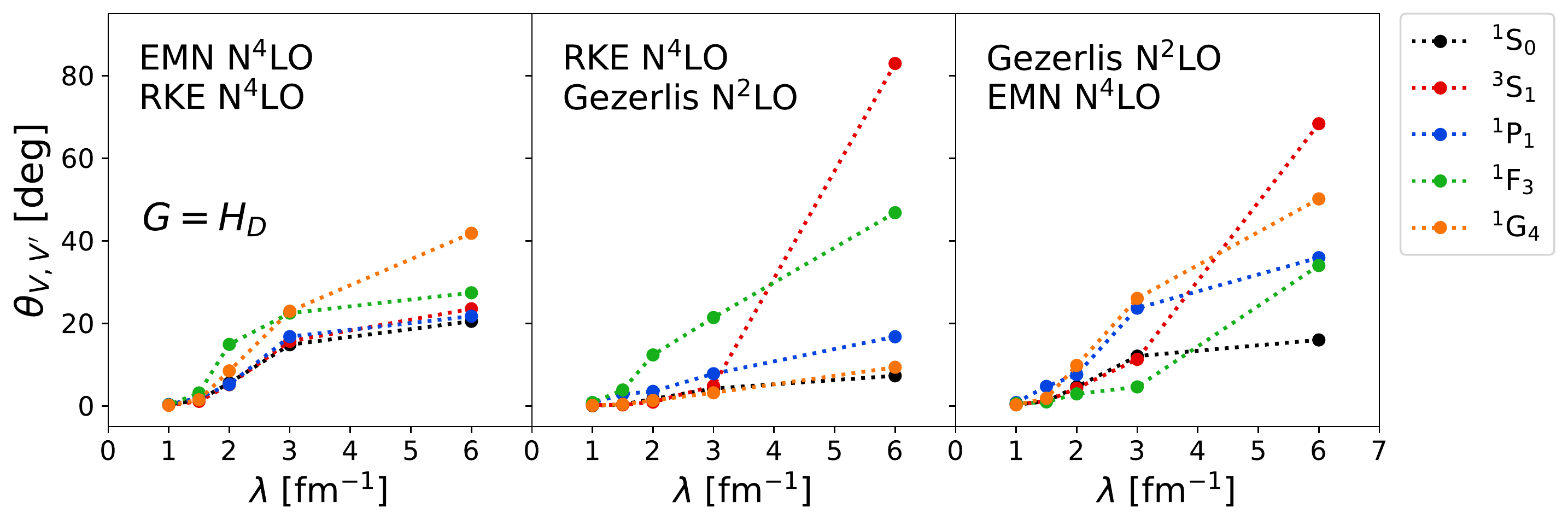}%
    \caption{SDT angle $\theta_{V,V'}$ between two SRG-evolved potentials for several partial wave channels, $^1$S$_0$ (black), $^3$S$_1$ (red), $^1$P$_1$ (blue), $^1$F$_3$ (green), and $^1$G$_4$ (orange), comparing EMN N$^4$LO 500 MeV, RKE N$^4$LO 450 MeV, and Gezerlis \textit{et al.}~N$^2$LO 1 fm.}
    \label{fig:universality_test_with_sdt_coeff_Wegner}
\end{figure*}

In Ref.~\cite{Dainton:2013axa}, it was found that shared long-distance physics (e.g., the common pion-exchange tail) plus phase equivalence up to some value of scattering momentum $k_0$ implies potential matrix element equivalence up to the same value $k_0$ in SRG-evolved potentials where $\lambda$, $\LambdaBD \leq k_0$ (see Figs.\ 1--4 in \cite{Dainton:2013axa}).
We verify the conclusion from~\cite{Dainton:2013axa} in the representative chiral potentials showing the $^3$S$_1$ channel as an example.
Figure~\ref{fig:potential_universality_3S1}(a) shows the NN phase shifts of EMN N$^4$LO 500 MeV, RKE N$^4$LO 450 MeV, and Gezerlis \textit{et al.}~N$^2$LO 1 fm potentials in the $^3$S$_1$ partial wave channel.
Figure~\ref{fig:potential_universality_3S1}(b) shows the diagonal and far off-diagonal matrix elements of the evolved potentials in the $^3$S$_1$ channel on the top and bottom row, respectively.
Band- and block-diagonal evolved potentials are shown on the same sub-plots indicated by solid and dashed lines, respectively, where the color indicates the potential.
The $^3$S$_1$ phase shifts agree to within 1\% for $k \leq 2$\,fm$^{-1}$, and as we see in Fig.~\ref{fig:potential_universality_3S1}(b), universality occurs once the potentials are SRG evolved past $2$\,fm$^{-1}$.
The matrix elements of the potentials all begin to collapse to the same line as $\lambda$ and $\LambdaBD$ decrease to the point of phase equivalence.
In this sense, we can think of the SRG evolution like an attractor; the potentials evolve in the same manner contingent on the SRG generator with a wide variety of starting points.
The two generators collapse the potential to a different form, because the induced contributions from SRG flow depend on how the potential is decoupled, that is, the choice in $G$.

\subsection{Quantifying universality}
\label{subsec:quantifying_universality}

Next, to quantify universality in the potentials, we calculate the Frobenius norm of the difference in potentials in Fig.~\ref{fig:universality_test_with_norm_Wegner}.
Here, we use $G=H_D$ as an example for several partial wave channels.
We evolve the three default potentials to $\lambda=6$, $3$, $2$, $1.5$, and $1$\,fm$^{-1}$.
To focus on the region of universality, we truncate each potential matrix up to the momentum value $\lambda$ and divide the difference by the average norm of the three truncated potentials.
(This prevents the norm from decreasing with lower $\lambda$ because the matrices become smaller in dimension due to truncation.)
The momentum value of phase equivalence occurs somewhere in the range of $1$--$2$\,fm$^{-1}$ for most of the channels included.
We see a sharp drop in the matrix norm at $\lambda$ near this range.

Notice that the $^3$S$_1$ channel differs significantly in comparing Gezerlis N$^2$LO to the other two potentials.
This is due to the $^3$S$_1$ being dominated by the contact force, where for Gezerlis N$^2$LO the regulator function is local, while it is  non-local for the other two potentials.
A similar difference is seen in the $^1$G$_4$ channel, which is dominated by pion-exchange, but now for EMN N$^4$LO compared to the other two.
Again, this is caused by the difference in regulator functions, where EMN N$^4$LO uses a non-local regulator and the other two a local regulator.
The difference in regulator functions between the various potentials affects the flow to universality in channels primarily affected by contact forces or pion exchange, but the difference is small and unnoticeable in the previous figures.

We can also use techniques from spectral distribution theory (SDT) to analyze universality~\cite{Johnson:2017uor}.
In SDT the expectation value of a potential is defined as
\begin{eqnarray}
    \label{eq:sdt_exp_value}
    \expval{V} = \frac{1}{N} \Tr V,
\end{eqnarray}
where $N$ is the dimension of the matrix $V$.
The inner product of two potentials $V$ and $V'$ is defined as,
\begin{eqnarray}
    \label{eq:sdt_inner_product}
    (V,V') &=& \expval{(V^{\dagger}-\expval{V^{\dagger}})(V'-\expval{V'})} \nonumber \\
    &=& \expval{V^{\dagger} V'} - \expval{V^{\dagger}}\expval{V'}.
\end{eqnarray}
We can now define the correlation coefficient $\zeta_{V,V'}$ which gives a measure of the ``similarity'' between the two potentials,
\begin{eqnarray}
    \label{eq:sdt_corr_coeff}
    \zeta_{V,V'} = \frac{(V,V')}{\sigma_V \sigma_{V'}},
\end{eqnarray}
where $\sigma_V$ is the positive square root of the variance,
\begin{eqnarray}
    \label{eq:sdt_variance}
    \sigma^2_V = (V,V) = \expval{V^2} - \expval{V}^2.
\end{eqnarray}
Geometrically, we can think of the potentials as two vectors with $\theta_{V,V'} \equiv \textrm{arccos}(\zeta_{V,V'})$ measuring the angle between them.
Further details of the relevant formulas in SDT can be found in Refs.~\cite{Sviratcheva:2007bp,Launey:2013ysa}.

Although these calculations have been used to quantify the differences in nuclear Hamiltonians, we provide calculations of $\theta_{V,V'}$ instead since the relative kinetic energy is the same in the three representative Hamiltonians.
Analogous to Fig.~\ref{fig:universality_test_with_norm_Wegner}, we show the angle between pairs of the potentials for the same values of $\lambda$ in Fig.~\ref{fig:universality_test_with_sdt_coeff_Wegner}.
Again, we make a truncation in the potential matrices up to the value of $\lambda$.
With SRG evolution, $\theta_{V,V'} \rightarrow 0$ corresponding to strong correlations between the compared potentials.
The differences in the behavior of the various channels as noted previously show in Fig.~\ref{fig:universality_test_with_sdt_coeff_Wegner} as well.

\subsection{Evolved wave functions and SRCs}
\label{subsec:wfs_and_SRCs}

\begin{figure}[t!]
    \includegraphics[clip,width=0.48\textwidth]{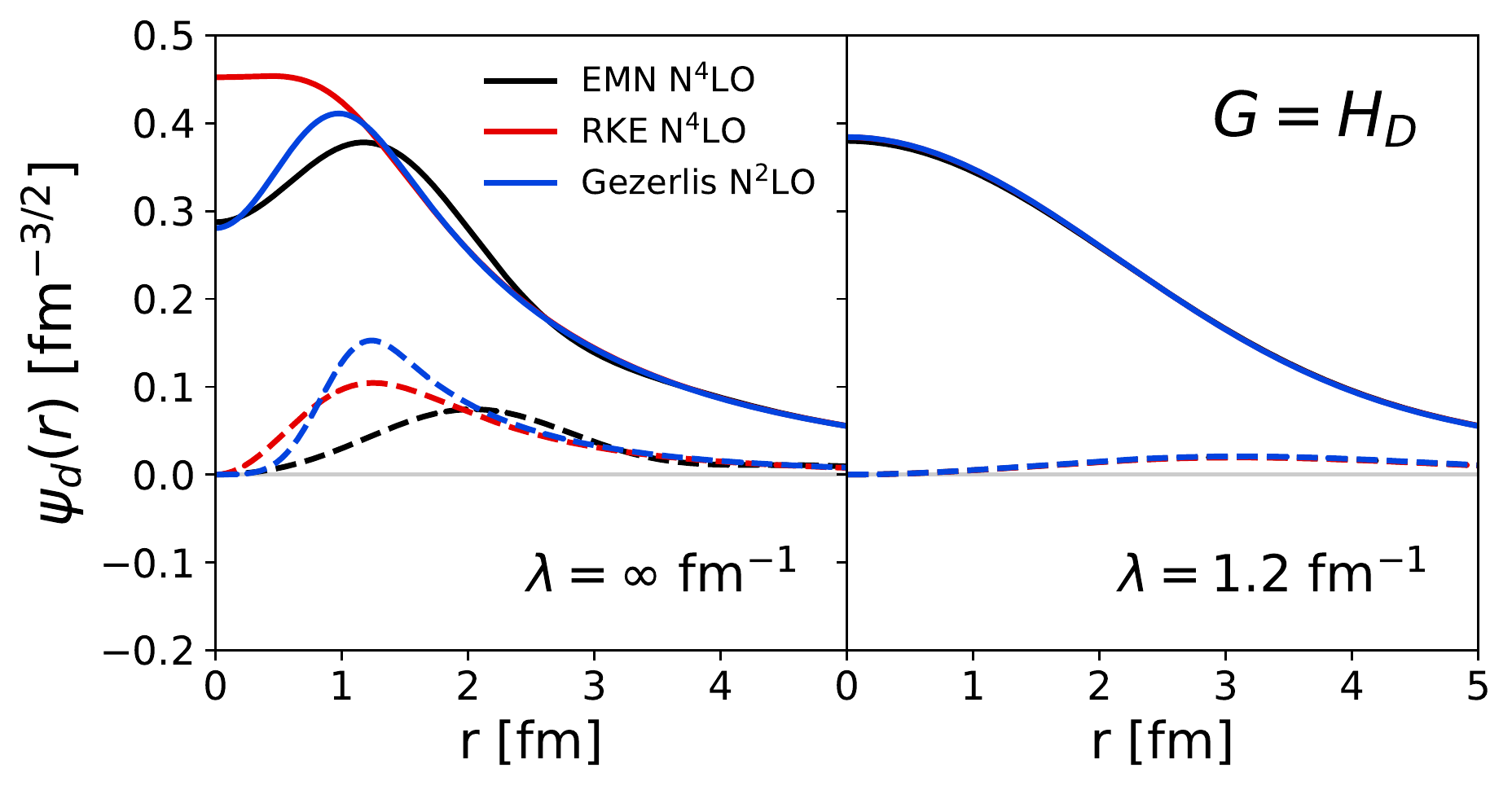}%
    \caption{Deuteron wave functions in coordinate space for the same three chiral potentials as in Fig.~\ref{fig:universality_test_with_sdt_coeff_Wegner} under band-diagonal SRG transformations with $G=H_D$ and $\lambda=1.2$\,fm$^{-1}$. The solid lines correspond to the S-states, and the dashed lines correspond to the D-states.}
    \label{fig:deuteron_wave_funcs_soft}
\end{figure}
\begin{figure}[tbh]
    \includegraphics[clip,width=0.48\textwidth]{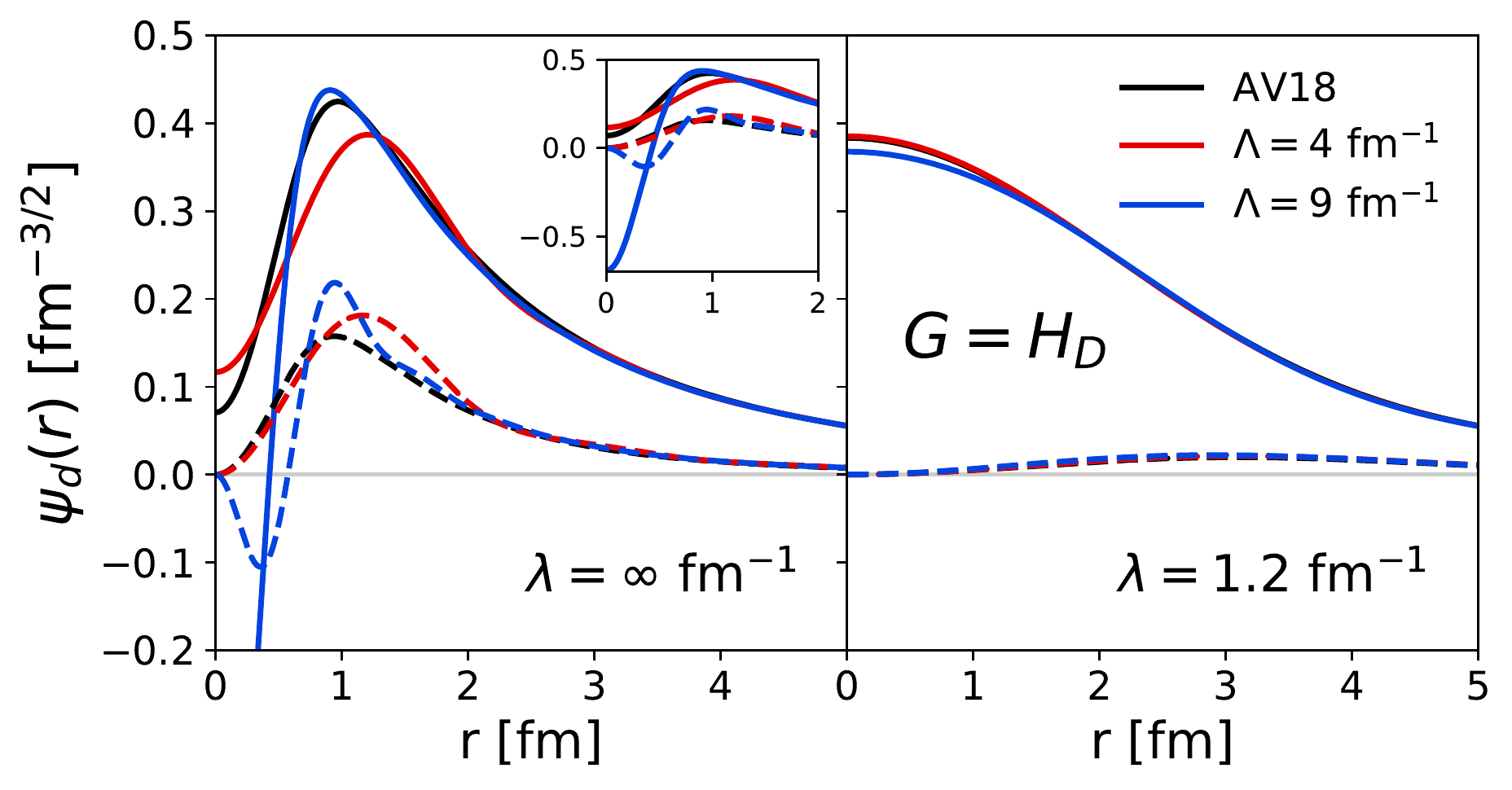}%
    \caption{Same as Fig.~\ref{fig:deuteron_wave_funcs_soft} but for the high-cutoff LO potentials with $\Lambda=4$ and $9$\,fm$^{-1}$ and AV18, and $\lambda=1.2$\,fm$^{-1}$. The inset plot on the left panel shows the initial wave functions zoomed out on the y-axis up to $r=2$ fm.}
    \label{fig:deuteron_wave_funcs_hard}
\end{figure}

Universality in the potentials for a given SRG generator is naturally reflected in the low-energy wave functions.
In Fig.~\ref{fig:deuteron_wave_funcs_soft} we show the initial and evolved deuteron wave functions in coordinate space for the three chiral potentials, where the solid lines correspond to the S-state components and the dashed lines to the D-state components.
The wave functions are evolved using a band-diagonal, $\lambda=1.2$\,fm$^{-1}$ transformation.
The short-distance part of the S-state differs initially but flows to the same form, while the initial D-state also differs and becomes suppressed after evolving.
This reflects the flow to universality in the low-momentum matrix elements of the potentials.
Despite the scheme dependence of the initial UV treatment, decoupling the low- and high-momentum physics means the states flow to the same wave function at low resolution.
Furthermore, the same low-resolution wave functions result for initial
deuteron wave functions with harder potentials such as Argonne $v_{18}$ (AV18) \cite{Wiringa:1994wb} and the LO high-cutoff potentials from the previous section, as seen in Fig.~\ref{fig:deuteron_wave_funcs_hard}.

Consider these results from the perspective of SRC phenomenology~\cite{Hen:2016kwk,Weiss:2016obx,Cruz-Torres:2019fum,Schmidt:2020kcl}.
In Fig.~\ref{fig:deuteron_wave_funcs_hard}, the dip at small $r$ in $\psi_d(r)$ in the initial S-states reflects a strong repulsive core in the initial potential (with the node for $\Lambda=9$\,fm$^{-1}$ because of the spurious deep-bound state for that potential) while the D-wave strength at short distance is from a strong tensor force.
These are the signatures of the SRC proton-neutron pair in the deuteron; there will be corresponding intermediate-momentum (D-state) and high-momentum (S-state) signatures in the momentum wave functions.
Note the qualitative similarity of AV18 and the chiral $\Lambda=4$\,fm$^{-1}$ wave functions, which demonstrates that even though the LO chiral potential is only adjusted to fit very low energy phase shifts, the same SRC structure is found because of the common iterated-pion-exchange and similar regularization scale in the respective Hamiltonians.
(This suggests that one is unlikely to explore fine details of the NN interaction from SRC physics.)
The higher momenta extend well beyond the chiral EFT breakdown scale of about 3\,fm$^{-1}$, where UV physics is incorrect.

The scale and scheme dependence of SRCs is manifest in these two figures.
But by shifting the resolution scale through SRG evolution, the SRC physics is dissolved as the deuteron state becomes decoupled from high-energy contributions.
All physical observables will be preserved with these uncorrelated wave functions if the corresponding operators are also SRG evolved.
The purely high-momentum contributions removed from the wave function are compensated in the evolved operator as smeared contact operators, as illustrated in the following section.
This reflects a natural factorization of the short-distance physics for low-energy states, which will be the same in all nuclei (with $^1$S$_0$ contributions as well for $A>2$).
This factorization accounts for the short-distance or high-momentum pair distributions for a \emph{fixed} high-resolution Hamiltonian being the same as well (so they are universal in a difference sense than we have been considering)~\cite{Anderson:2010aq,Bogner:2012zm}.

The flow to universality in the wave functions for a well-specified SRG scheme suggests that the lower resolution scales for nuclear structure from soft potentials and the shell model can be matched by a well-specified reaction operator structure~\cite{More:2017syr}.  
The S-state versus D-state probabilities of the deuteron can be viewed as spectroscopic factors for single-particle strengths~\cite{Furnstahl:2010wd}.
If one analyzed scattering from the deuteron S-state with a high-resolution reaction model but the low-resolution wave function, the reduced D-state component would lead one to conclude that the ratio of experiment to theory cross sections was less than one.
This is the analog of what is found in knock-out experiments analyzed with an eikonal reaction model and shell model wave functions~\cite{Tostevin:2014usa}.
The flow to universal structure may provide a controlled resolution of these discrepancies.

In summary, we have examined the flow to universality of several recently developed \chiEFT\ potentials.
We verified that the general conclusions of Ref.~\cite{Dainton:2013axa} still hold, namely that potential matrix elements collapse to similar values in regions of phase equivalence. 
We quantified this collapse using both Frobenius norm and the SDT angle $\theta_{V,V'}$, which highlight the differences in the three representative potentials from the regulator functions.
Lastly, we illustrated the consequence of this universality for low-energy wave functions of the potentials by applying transformations to the deuteron.

\section{Evolution of other operators}
\label{sec:evolution_other_operators}

\begin{figure*}[t!]
	\includegraphics[clip,width=0.85\textwidth]{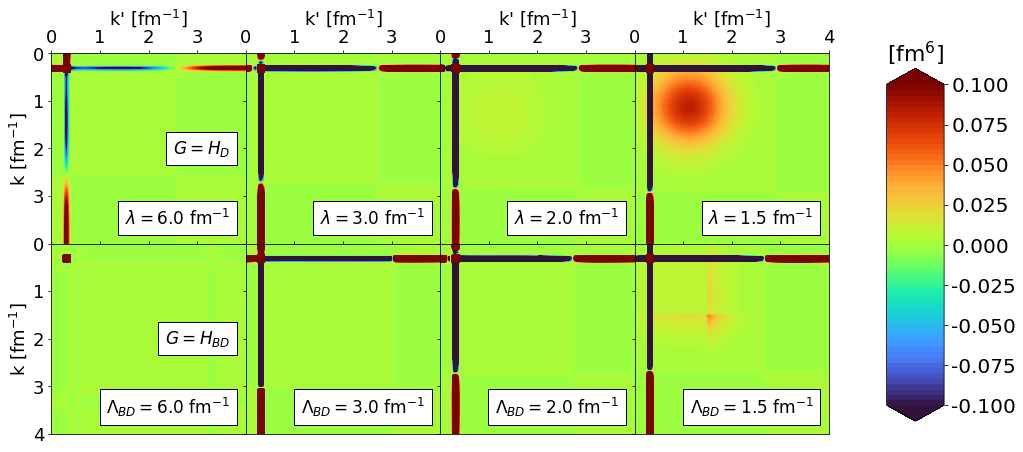}%
	\caption{Momentum projection operator $\mel{k}{\ataq}{k'}$ for $q=0.3$\,fm$^{-1}$ under SRG transformations using the RKE N$^4$LO 450 MeV potential, evolving with Wegner ($H_D$) and block-diagonal ($H_{BD}$) generators in the $^3$S$_1$ channel. The SRG flow parameter $\lambda$ is varied for $G = H_D$ evolution and fixed at $\lambda=1$\,fm$^{-1}$ for $G = H_{BD}$ evolution. The decoupling scale for $G = H_{BD}$ is $\LambdaBD$.}
	\label{fig:momentum_projection_contours_q0,30_3S1_RKE}

\medskip\smallskip

	\includegraphics[clip,width=0.85\textwidth]{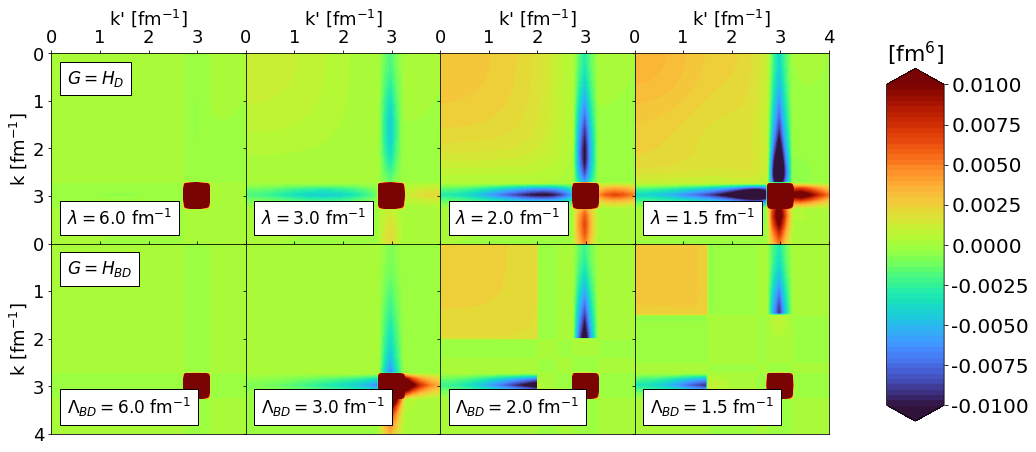}%
	\caption{Same as Fig.~\ref{fig:momentum_projection_contours_q0,30_3S1_RKE} but for $q=3$\,fm$^{-1}$.}
	\label{fig:momentum_projection_contours_q3,00_3S1_RKE}
\end{figure*}
%

\subsection{SRG for representative operators}
\label{subsec:representative_ops}

In this section, we analyze SRG operator evolution using the radius squared operator $r^2$ and the momentum projection operator $\ataq$, where $q$ is the relative momentum.
These serve as examples of long-distance operators ($r^2$) and low- and high-momentum operators (by specifying different $q$ in $\ataq$). 
We look to whether the observations on universality for SRG-evolved Hamiltonians can be generalized to universality for \emph{any} SRG-evolved operator and contrast the evolution for different generators.
In evolving these operators, we build the SRG unitary transformations explicitly using the eigenvectors of the bare and evolved Hamiltonians, that is,
\begin{eqnarray}
	\label{eq:unitary_transformation}
	U(s) = \sum_{\alpha=1}^{N} \ket{\psi_{\alpha}(s)} \bra{\psi_{\alpha}(0)},
\end{eqnarray}
where $\alpha$ indexes the states of the Hamiltonian.
Then to evolve the operator, we apply $U(s)$ as in Eq.~\eqref{eq:srg_operator}.

We start by considering the relative momentum projection operator,
$\ataq$, which works in a very simple way in the two-body system.
The expectation value of $\ataq$ in some state $\ket{\psi}$ gives the momentum distribution evaluated at $q$, that is, $\mel{\psi}{\ataq}{\psi}=\abs{\psi(q)}^2$.
Hence, the $k$, $k'$ matrix element of this operator is proportional to two delta functions: $\delta(k-q) \delta(k'-q)$.
In the simplest discretization, this corresponds to a matrix of zeros at every point in $k$ and $k'$ except where $k=k'=q$, which makes the SRG-induced contributions quite clear.
More generally, we can use smeared delta functions with non-zero entries, appropriately weighted to integrate to one, for the matrix elements near $k=k'=q$. 
In Figs.~\ref{fig:momentum_projection_contours_q0,30_3S1_RKE} and \ref{fig:momentum_projection_contours_q3,00_3S1_RKE} we show two different sets of SRG-evolved momentum projection operators (for $q=0.3$ and $3$\,fm$^{-1}$) with the Wegner and block-diagonal generators using a slightly smeared operator.
For large $\lambda$ and $\LambdaBD$ values in both figures, we see the initial regularized delta functions as a dark red dot where $k=k'=q$, which persists with SRG evolution.
(See Figs. 3 and 4 in Ref.~\cite{Anderson:2010aq} for similar visualizations with the simplest discretization.)

In Fig.~\ref{fig:momentum_projection_contours_q0,30_3S1_RKE} where $q=0.3$\,fm$^{-1}$, the most evident induced contributions are non-zero bands for $k=q$ or $k'=q$ and then smooth induced contributions eventually become visible at $k,k'<2$\,fm$^{-1}$.
These features are independent of the smearing of the delta function and matrix elements in the deuteron are the same up to small discretization artifacts.
We can understand the bands by taking one infinitestimal step $\Delta s$ in the SRG evolution in Eq.~\eqref{eq:srg_flow} and taking $k,k'$ matrix elements,
\begin{align}
    \mel{k}{\Delta \ataq(s)}{k'} &= \mel{k}{[\eta, \ataq(0)]}{k'} \Delta s.
\end{align}
After inserting an intermediate integration, $\ataq(0)$ will evaluate to two (smeared) delta functions, one of which survives each term as $\delta(k-q)$ or $\delta(k'-q)$.
These delta function contributions persist throughout the evolution and therefore show up as (smeared) bands.
For the Wegner generator the low-momentum induced contributions are much larger than in the case of the block-diagonal generator. 
This reflects that the Hamiltonian is being modified more at high momentum by the band-diagonal evolution.
Consequently the block-diagonal transformation roughly keeps the same low-$k$ wave function (assuming a low-energy state), therefore a low-momentum operator will change less under block-diagonal transformations.
This is analyzed further in Sec.~\ref{subsec:connecting}.

Figure~\ref{fig:momentum_projection_contours_q3,00_3S1_RKE} shows SRG evolution of $\mel{k}{\ataq}{k'}$ again but for $q=3$\,fm$^{-1}$.
The band-diagonal SRG transformation induces low-momentum contributions where the initial operator was entirely zero.
A similar change happens for the block-diagonal transformation except the induced contributions at low momentum sharply drop to zero at the block-diagonal cutoff $\LambdaBD$.
The smooth low-momentum contributions are what is expected from an EFT perspective, as they can be expanded as regulated (smeared) contact operators that absorb the high-momentum contributions to low-energy states that are decoupled by the evolution.
These features are independent of the mesh and the discretization of the delta functions.
This is an example of an operator product expansion factorization~\cite{Anderson:2010aq,Bogner:2012zm}, which is reviewed in Sec.~\ref{subsec:factorization}.

\begin{figure}[t!]
    \includegraphics[clip,width=0.48\textwidth]{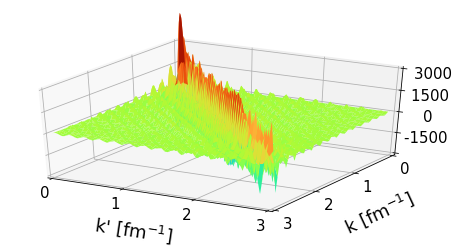}%
    \caption{Visualization of the $r^2$ operator in momentum space, $\langle k | r^2 | k'\rangle$, regulated only by the coordinate and momentum meshes.
    The integration factors of $k$ and $k'$ are included.}
    \label{fig:r2_unregulated}
    
\medskip\smallskip

    \includegraphics[clip,width=0.48\textwidth]{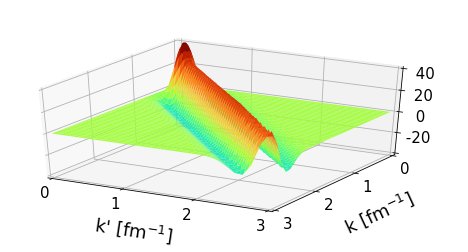}%
    \caption{Same as Fig.~\ref{fig:r2_unregulated} but with a coordinate-space regulator function $e^{-r^2/a^2}$ where $a=6$ fm.}
    \label{fig:r2_regulated}
\end{figure}
\begin{figure*}[htb]
    \includegraphics[clip,width=0.9\textwidth]{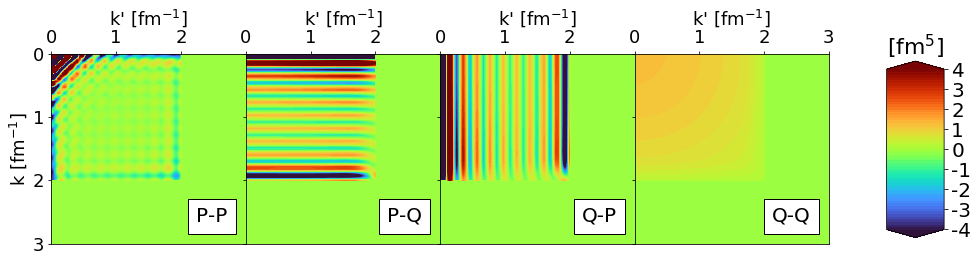}%
    \caption{SRG contributions to $P r^2 P$ in momentum space, splitting $PU(\LambdaBD)r^2(\infty)U^{\dagger}(\LambdaBD)P$ into four components as in Eq.~\eqref{eq:r2_PP_contributions}. For the $\PP$ contribution, the unevolved operator $P r^2 P$ is subtracted out. We apply a block-diagonal transformation from RKE N$^4$LO 450 MeV in the $^3$S$_1$ channel with $\LambdaBD=2$\,fm$^{-1}$.}
    \label{fig:r2_srg_changes}
\end{figure*}

Next, we consider the $r^2$ operator, relying on Ref.~\cite{Anderson:2010aq} for formulas and some basic results.
In the absence of an explicit regulator for large $r$, the meshes used to create the $r^2$ matrix in coordinate space and then Fourier transform to momentum will act as regulators. 
Visualizations of the bare $r^2$ operator will then be highly mesh dependent, even though expectation values of $r^2$ will be stable for sufficiently large cutoff in $r$ or small mesh spacing in $k$.
With this in mind, we show visualizations of $\langle k | r^2 | k'\rangle$ in Figs.~\ref{fig:r2_unregulated} and \ref{fig:r2_regulated}, where the latter has an added regularization to illustrate the strong regulator dependence.
(The $r^2$ operator in Figs.~\ref{fig:r2_unregulated} and \ref{fig:r2_regulated} include integration factors $k$ and $k'$ such that evaluating $\mel{\psi}{r^2}{\psi}$ in momentum space with wave functions equipped with additional factors $k$ or $k'$ will give the correct integration.)
As evident in both figures, there is strength near the diagonal for all $k$, so the contribution to the $r^2$ expectation value for a particular state will be dictated by its momentum wave function.

\begin{figure*}[tbh]
	\subfloat[]{%
	\includegraphics[clip,width=0.35\textwidth]{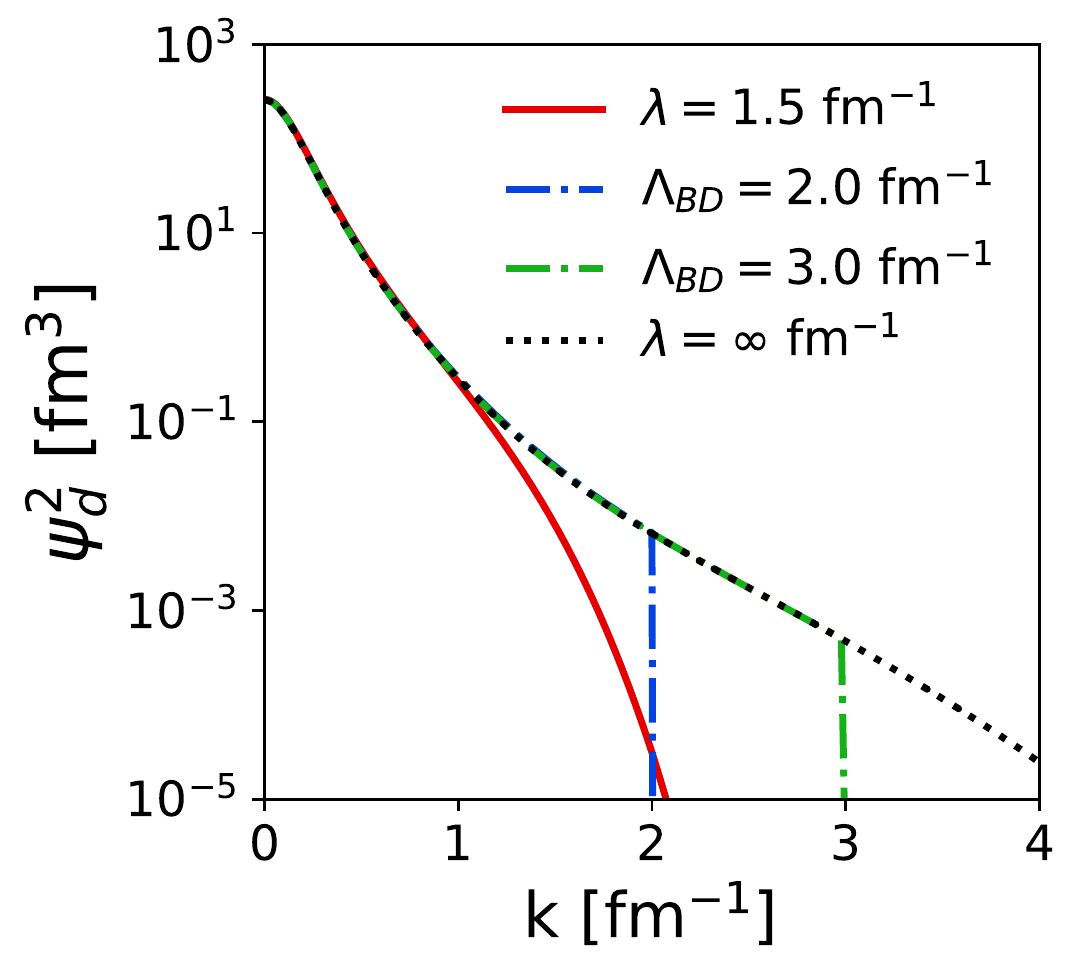}%
	}
	\quad
	\subfloat[]{%
	\includegraphics[clip,width=0.35\textwidth]{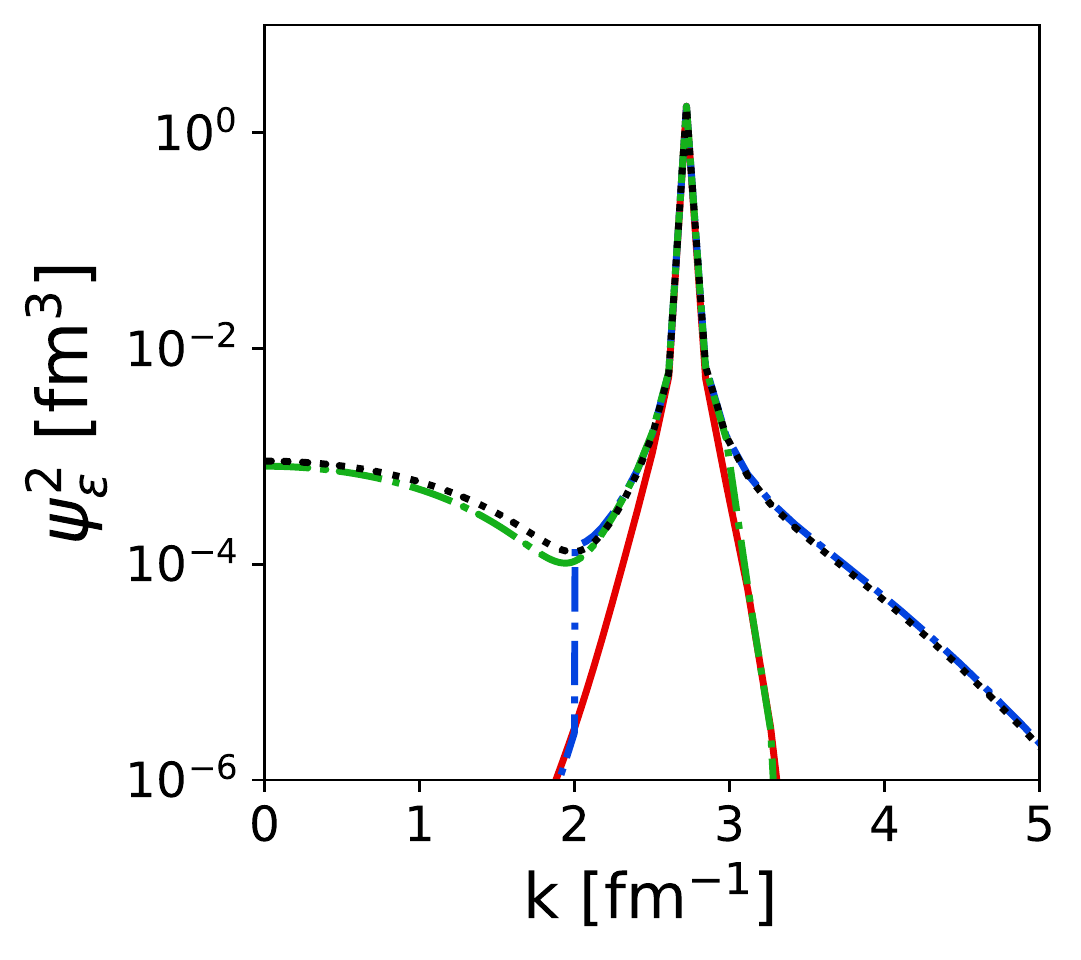}%
	}
	\caption{Momentum distributions from deuteron (a) and a high-energy state (b) with the RKE N$^4$LO 450 MeV potential. Here, we compare SRG-evolved distributions, $G=H_D$ (red solid), $G=H_{BD}$ at $\LambdaBD=2$\,fm$^{-1}$ (blue dash-dotted), and $G=H_{BD}$ at $\LambdaBD=3$\,fm$^{-1}$ (green dash-dotted) each with $\lambda=1.5$\,fm$^{-1}$, to the initial distribution (black dotted). Also, $\varepsilon \approx 300$ MeV for the high-energy state in (b).}
	\label{fig:momentum_distributions_RKE}
\end{figure*}

The SRG evolution of the regulated $r^2$ operator is barely noticeable in contour plots (e.g., see Ref.~\cite{Anderson:2010aq}), so we focus instead on the induced changes in the operator when evolving to low momentum.
We use block-diagonal evolution for clarity and split the contribution to $r^2$ according to its origin in different blocks of momentum space defined by $P = \theta (\LambdaBD - k)$ and $Q = \theta(k - \LambdaBD)$ projection operators.
In particular, the four contributions to the evolved $r^2$ operator (designated $r^2(\LambdaBD$)) in the low-momentum block is decomposed as:
\begin{align}
    \label{eq:r2_PP_contributions}
    P r^2(\LambdaBD) P &= P U(\LambdaBD) P r^2(\infty) P U^{\dagger}(\LambdaBD) P \nonumber \\
    & \null + P U(\LambdaBD) P r^2(\infty) Q U^{\dagger}(\LambdaBD) P \nonumber \\
    & \null + P U(\LambdaBD) Q r^2(\infty) P U^{\dagger}(\LambdaBD) P \nonumber \\
    & \null + P U(\LambdaBD) Q r^2(\infty) Q U^{\dagger}(\LambdaBD) P.
\end{align}
In Fig.~\ref{fig:r2_srg_changes} we show a representative set of these contributions for the RKE N$^4$LO 450\,MeV potential, labeled by their origin before the block-diagonal unitary transformations.

The $\QQ$ panel in Fig.~\ref{fig:r2_srg_changes} is very similar to the corresponding $\PP$ block for the evolved high-momentum $\ataq$ shown in Fig.~\ref{fig:momentum_projection_contours_q3,00_3S1_RKE}.
This is not a coincidence: these smooth low-momentum contributions have the same origin and same understanding from EFT and OPE factorization.
The $\PQ$ block is roughly constant in $k'$ for the same reason, while the $k$ dependence is associated with the bare operator, and therefore with the regularization (here as in Fig.~\ref{fig:r2_unregulated}), as is the full contribution in the $\PP$ block (likewise for the $\QP$ block, swapping $k'$ and $k$).
The implications for matrix elements in the deuteron are given in the next section.
We discuss how the behavior in each panel of Fig.~\ref{fig:r2_srg_changes} follows from factorization in Sec.~\ref{subsec:factorization}.
The decomposition for other potentials or other choices for $\LambdaBD$ is qualitatively similar, with the $\QQ$ contribution scaling with the hardness of the interaction, which reflects the extent of initial high-momentum components (i.e., the short-range correlations).

\begin{figure*}[tbh]
	\includegraphics[clip,width=0.85\textwidth]{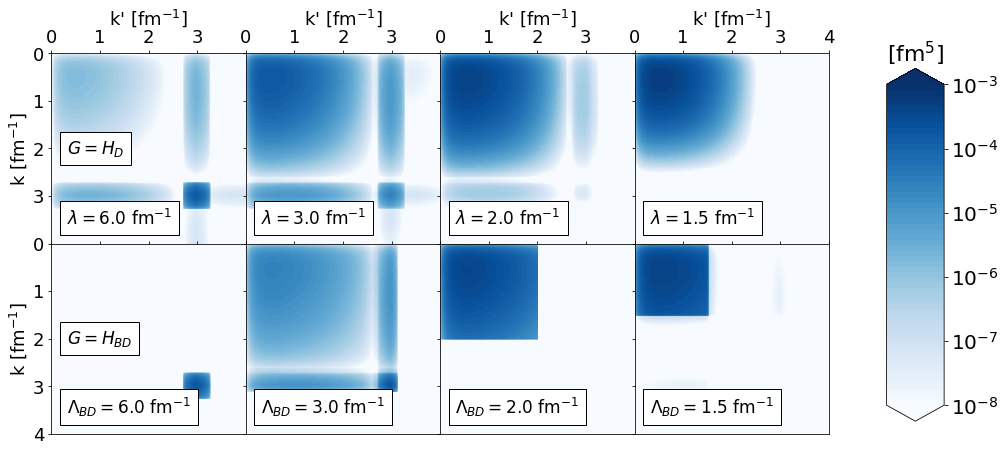}%
	\caption{Integrand of $\mel{\psi_d}{\ataq}{\psi_d}$ in momentum space where $\psi_d$ is the deuteron wave function and $q=3$\,fm$^{-1}$. Here, we SRG-evolve the operator and wave function where each successive column indicates further evolution under the Wegner and block-diagonal generators with the RKE N$^4$LO 450 MeV potential. We vary the SRG flow parameter $\lambda$ for Wegner evolution and fix it at $\lambda=1$\,fm$^{-1}$ for block-diagonal evolution. The decoupling scale in the block-diagonal generator is denoted by $\LambdaBD$.}
	\label{fig:momentum_projection_integrand_contours_q3,00_RKE}
\end{figure*}
\begin{figure*}[tbh]
	\includegraphics[clip,width=0.85\textwidth]{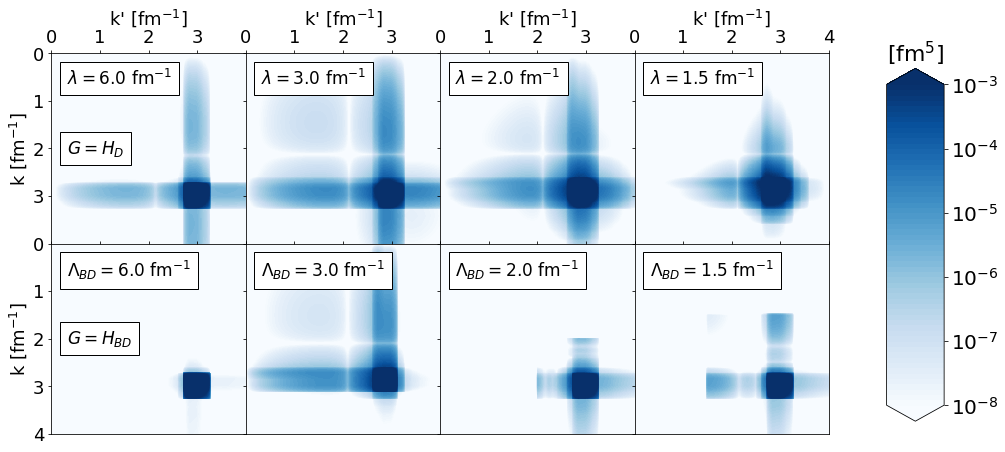}%
	\caption{Same as Fig.~\ref{fig:momentum_projection_integrand_contours_q3,00_RKE} but with a high-energy state $\psi_{\varepsilon}$ where $\varepsilon \approx 300$ MeV.}
	\label{fig:momentum_projection_integrand_contours_q3,00_eps300_RKE}
\end{figure*}
%

\subsection{Connecting to wave function evolution}
\label{subsec:connecting}

\begin{table*}[htb]
	\caption{SRG contributions to $\mel{\psi_d}{P r^2 P}{\psi_d}$ splitting $PU(\LambdaBD)r^2(\infty)U^{\dagger}(\LambdaBD)P$ into four components as in Eq.~\eqref{eq:r2_PP_contributions} where $\PQ$ and $\QP$ are combined.
	For the $\PP$ contribution, the unevolved $\mel{\psi_d}{P r^2 P}{\psi_d}$ value is subtracted.
	We apply block-diagonal transformations from RKE N$^4$LO 450 MeV and AV18 in the $^3$S$_1$ channel with $\LambdaBD=2\,$fm$^{-1}$. For comparison, we also show results for $\ataq$ with $q = 3\,\mbox{fm}^{-1}$.}
	\label{tab:evolved_r2_contributions}
	\begin{ruledtabular}
		\begin{tabular}{c||ccc||ccc}
		   & \multicolumn{3}{c||}{$\mel{\psi_d}{P r^2 P}{\psi_d}$ [fm$^2$]} &
		   \multicolumn{3}{c}{$\mel{\psi_d}{P \ataq P}{\psi_d}$ [fm$^3$]} \\
      	\mystrut	Potential &  $\PP$ & $\PQ+\QP$ & $\QQ$ & $\PP$ & $\PQ+\QP$ & $\QQ$\\
			\colrule
      			RKE N$^4$LO 450\,MeV & $\num{-2.90e-02}$ & $\num{-3.22e-01}$ & $\num{1.66e-01}$ &
      			$\num{0.0}$ & $\num{0.0}$ & $\num{5.05e-04}$
                    \mystrut\\
      			AV18 & 
      			\num{-4.83e-02} & \num{-4.33e-01} & \num{2.33e-01}
      			& $\num{0.0}$ & $\num{0.0}$  & \num{1.61e-03} \mystrut
		\end{tabular}
  	\end{ruledtabular}
\end{table*}
\begin{figure}[htb]
    \includegraphics[clip,width=0.4\textwidth]{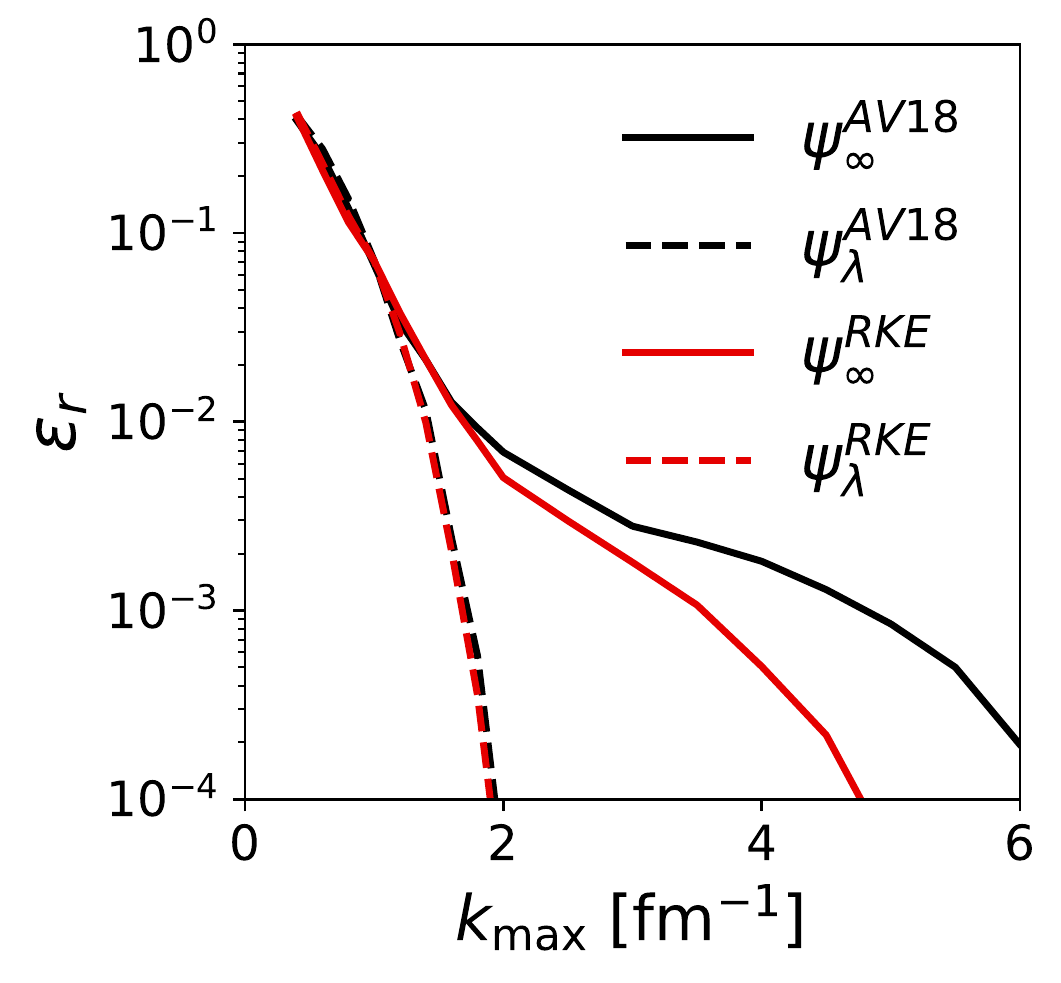}%
    \caption{Relative error of the deuteron rms radius from AV18 (black) and RKE N$^4$LO $450$ MeV (red) truncating the momentum-space calculation $\mel{\psi_d}{r^2}{\psi_d}$ at $k_{\rm max}$. Solid lines indicate the fully unevolved calculation and dashed lines indicate the SRG-evolved calculation where $\lambda=1.5$\,fm$^{-1}$.}
    \label{fig:r_convergence_with_kmax}
\end{figure}

SRG transformations are unitary, meaning that the matrix elements of the evolved operator are preserved.
Therefore, the changes in the operator must be accounted for in the evolved wave functions.
We can examine the evolved wave functions to understand the differences in band- and block-diagonal evolution of operators.

Figure~\ref{fig:momentum_distributions_RKE} shows initial and evolved momentum distributions for the deuteron and a high-energy state at $\varepsilon \approx 300$ MeV using the RKE N$^4$LO 450 MeV potential.
For the deuteron, the strength of the wave function is shifted to lower momentum with the band-diagonal generator.
With the block-diagonal generators, the wave function is nearly the same up to the value of the cutoff $\LambdaBD$.
For the high-energy state, the band-diagonal generator keeps the strength of the wave function near the spike at $k \approx 2.7$\,fm$^{-1}$.
However, in the case of the block-diagonal generator, the wave function changes in different ways depending on the cutoff $\LambdaBD$.
For $\LambdaBD=2$\,fm$^{-1}$, we see the evolved distribution roughly matching the initial one for $k>\LambdaBD$, and vice versa for $\LambdaBD=3$\,fm$^{-1}$. 
Recall that in block-diagonal SRG decoupling the Hamiltonian is split into a low-momentum sub-block, $P H P$, and a high-momentum sub-block, $Q H Q$.
When $\LambdaBD=2$\,fm$^{-1}$, the $\varepsilon \approx 300$ MeV state is contained in $Q H Q$, whereas for $\LambdaBD=3$\,fm$^{-1}$ it is contained in $P H P$.
Note that the deuteron, being the lowest energy state, is also contained in $P H P$, which is consistent with what is seen in Fig.~\ref{fig:momentum_distributions_RKE}(a).
A block-diagonal-evolved wave function remains approximately unchanged in the sub-block where the state resides with the rest of the wave function dropping to zero.

Generally speaking, SRG transformations change operators based on how the transformations change the wave functions, which depends on the type of decoupling.
Consider the momentum projection operator with $q=3$\,fm$^{-1}$ and a block-diagonal transformation with $\LambdaBD=2$\,fm$^{-1}$.
We see the evolved wave functions for the deuteron and the high-energy state are opposite in the sense that the evolved deuteron wave function matches the initial wave function for $k<\LambdaBD$ and the high-energy state wave function matches for $k>\LambdaBD$.

We can use the momentum projection operator $\ataq$ to understand the contrasting behavior in the wave functions.
With $\mel{\psi(0)}{\ataq(0)}{\psi(0)} = \mel{\psi(s)}{\ataq(s)}{\psi(s)}$ from unitarity, how does the evolved projection operator for $q=3$\,fm$^{-1}$ and $\LambdaBD=2$\,fm$^{-1}$ make sense given these changes to the example wave functions?
For the deuteron wave function, the expectation value takes strength from the induced low-momentum contributions in the evolved operator where the evolved deuteron wave function is strongest ($k<2$\,fm$^{-1}$).
For the high-energy state, the expectation value depends more on the remnants of the delta functions from the initial operator because the strength of the wave function is at high momentum.
In each case, the expectation value remains the same.

\begin{figure*}[tbh]
	\includegraphics[clip,width=0.85\textwidth]{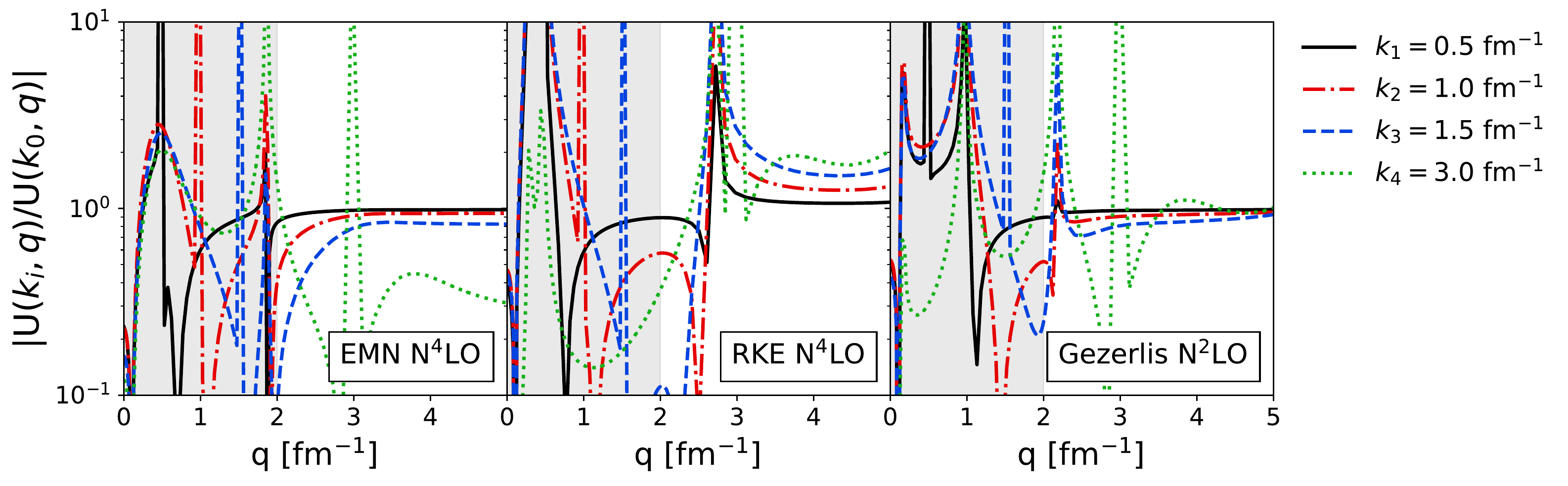}%
	\caption{Numerical tests of factorization of the unitary transformation  by plotting ratios of $|U(k,q)|$ in the $^3$S$_1$ channel as a function of $q$ for fixed $k=k_0$ in the denominator and several values of $k_i$ in the numerator.
    Plateaus in $q$ indicate factorization $U(k,q) \approx \Klo(k)\Khi(q)$, which is expected for $q\gg\lambda$ (outside shaded box) \emph{and} $k < \lambda$. The unitary transformations are generated in each panel for a different chiral EFT potential, all evolved to $\lambda = 2\,\mbox{fm}^{-1}$ with $G = H_D$.}
	\label{fig:factorization_ratios}
\end{figure*}

The SRG does not decouple \emph{every} operator in the sense that it decouples matrix elements as in the Hamiltonian, but instead reflects the changes made to the wave functions.
In Figs.~\ref{fig:momentum_projection_integrand_contours_q3,00_RKE} and \ref{fig:momentum_projection_integrand_contours_q3,00_eps300_RKE} we show the evolution of the integrand of the expectation value $\mel{\psi}{\ataq}{\psi}$ for deuteron and a high-energy state $\varepsilon \approx 300$ MeV, respectively, where $q=3$\,fm$^{-1}$.
Both the wave function and operator are SRG evolved so the total strength is preserved.

In Fig.~\ref{fig:momentum_projection_integrand_contours_q3,00_RKE}, the SRG transformations shift the strength in the integrand to lower momentum, matching the changes in the SRG-evolved deuteron wave function.
The band-diagonal transformation in the top row smoothly approaches lower and lower momentum, eliminating the delta functions, while the block-diagonal shows similar behavior but roughly sets an upper limit $\LambdaBD$ on the intermediate integrations.
The low-momentum contributions in the expectation value are relatively constant as the fall off of the wave function $\psi(k)$ largely cancels out with the integration factors $k^2$ and $k'^2$.
In Fig.~\ref{fig:momentum_projection_integrand_contours_q3,00_eps300_RKE}, band-diagonal evolution locally decouples the expectation value at high momentum.
Block-diagonal evolution sharply isolates the expectation value to low- or high-momentum sub-blocks depending on the value of $\LambdaBD$.
In the first two columns with block-diagonal decoupling, the expectation value resides in the low-momentum sub-block, but then switches to the high-momentum sub-block in the last two columns with lower $\LambdaBD$.

The expectation value $\mel{\psi_d}{r^2}{\psi_d}$ shows little variation with SRG evolution as the strength of initial operator resides predominantly at low momentum.
Thus, softening the high-momentum tail of the deuteron wave function leads to only a small change in the expectation value with $r^2$.
In Fig.~\ref{fig:r_convergence_with_kmax} the contributions from different regions in $k$ to the unevolved deuteron matrix element of $r^2$ are shown for the AV18 and RKE N$^4$LO 450\,MeV potentials by plotting the relative error made by integrating only up to $k_{\text{max}}$.
Only about 1\% of the expectation value comes from above 2\,fm$^{-1}$ in the initial wave function for either potential and there is negligible contribution above 2\,fm$^{-1}$ once they are evolved to $\lambda=1.5\,\mbox{fm}^{-1}$.
Table~\ref{tab:evolved_r2_contributions} shows the contributions from the blocks in Fig.~\ref{fig:r2_srg_changes} to the deuteron.
In contrast to the entries for $\ataq$ with $q = 3\,\mbox{fm}^{-1}$, which come entirely from the high-momentum \QQ\ block, the four blocks each contribute to the induced $r^2$ expectation values.
This implies that attempts to estimate the high-resolution SRC contribution to the low-resolution radius in schematic models, as in Ref.~\cite{Miller:2018mfb}, are rather subtle.
We have checked that contributions to $\mel{\psi_d}{P r^2P }{\psi_d}$ with transformations from other potentials are consistent with the results in Table~\ref{tab:evolved_r2_contributions}.

\subsection{Factorization}
\label{subsec:factorization}

In Refs.~\cite{Bogner:2007jb,Anderson:2010aq} it was shown that when there is a scale separation in its momentum arguments, the SRG unitary transformation should factorize into separate functions of low and high momentum, that is, $U(k,q) \rightarrow \Klo(k)\Khi(q)$ for $k < \lambda \ll q$.%
\footnote{In Ref.~\cite{Anderson:2010aq}, $\Klo$ was denoted $K$ and $\Khi$ was denoted $Q$. We switch notation here to avoid confusion with the projection operator $Q$.}
This is expected from general considerations of the operator product expansion~\cite{Anderson:2010aq,Bogner:2012zm}.
A test of factorization for three chiral EFT potentials is shown in Fig.~\ref{fig:factorization_ratios} by plotting the ratio $|U(k_i,q)/U(k_0,q)|$ versus $q$ with $k_0=0.1\,\mbox{fm}^{-1}$ for several different $k_i$. 
In general this ratio should vary widely with $q$ but should reduce to $|\Klo(k_i)/\Klo(k_0)|$, which is independent of $q$, when the conditions for factorization are satisfied.
This is validated in the figure as plateaus of the $U$ ratio in the expected region in $q$ for $k_i < \lambda$.
Furthermore, these plateaus are close to one and vary slowly with $k_i$, so the same is true of $\Klo(k)$.
We show only the $^3$S$_1$ channel in Fig.~\ref{fig:factorization_ratios} but have verified factorization in other channels as well.

Consider the consequences of this factorization for block-diagonal SRG evolution of an unevolved high-momentum operator, which we define as one with support only in the $\QQ$ block as in Sec.~\ref{subsec:representative_ops}:
\begin{align}
    [O_Q]_\infty = Q [O_Q]_\infty Q.
\end{align}
This includes $\ataq$ for $q$ in $Q$ and $Qr^2(\infty)Q$ in \eqref{eq:r2_PP_contributions}.
In the low-momentum block, the evolved operator becomes
\begin{align}
    P [O_Q]_\LambdaBD P &=
      P U(\LambdaBD)Q[O_Q]_\infty Q U^\dagger(\LambdaBD) P \notag \\
      &\approx P \Klo \bigl[\Khi [O_Q]_\infty \Khi\bigr] \Klo P
\end{align}
or, for $k,k'$ in the $\PP$ block,
\begin{align} \label{eq:factored_op}
   & \mel{k}{[O_Q]_\LambdaBD}{k'} \approx 
        \Klo(k) \Klo(k') \notag \\
  &  \quad \times
     \biggl[
      \int_{\LambdaBD}^{\infty} d\tilde{q}' 
      \int_{\LambdaBD}^{\infty} d\tilde{q}''\, \Khi(q') [O_Q]_\infty(q',q'') \Khi(q'')
     \biggr],
\end{align}
where $d\tilde{k} \equiv \frac{2}{\pi} k^2 dk$.
As all of the $k,k'$ dependence comes from the smooth functions $\Klo(k)\Klo(k')$, this universal result directly explains the particular cases of the $\PP$ block behavior in Fig.~\ref{fig:momentum_projection_contours_q3,00_3S1_RKE} and the $\QQ$ panel of Fig.~\ref{fig:r2_srg_changes}.
The same analysis goes through with the band-diagonal SRG, but with $\LambdaBD \rightarrow \lambda$ and the boundaries not so sharply defined.

Note that in the non-$\QQ$ panels of Fig.~\ref{fig:r2_srg_changes} at least one of the two SRG transformations does not factorize and is fully in the low-momentum $P$ sub-space.
For instance, in the $\PQ$ panel, we have one integral over low momentum and another over high momentum.
This yields oscillatory behavior in $k$ from the form of the bare $r^2$ operator (which is scheme dependent) and little variation in $k'$ because the factorized function $\Klo(k')$ is constant at leading order.

\begin{figure*}[tbh]
	\includegraphics[clip,width=0.75\textwidth]{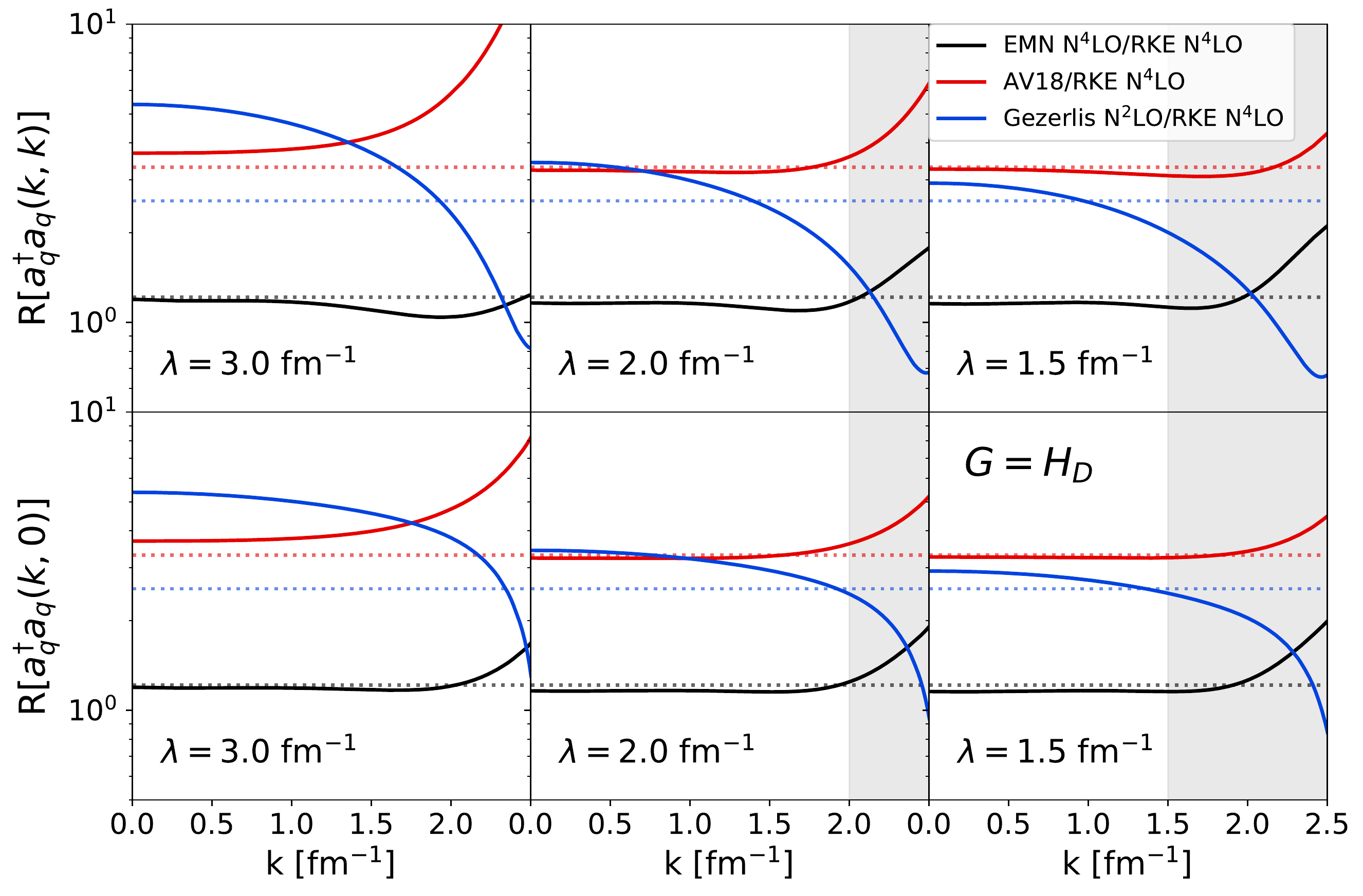}%
	\caption{Ratio of evolved matrix elements $R[\ataq(k,k)]$ (diagonal) and $R[\ataq(k,0)]$ (full off-diagonal) as defined in \eqref{eq:momentum_proj_matrix_elements_ratio} in the $^3$S$_1$ channel for $q = 3\,\mbox{fm}^{-1}$.
	Three potentials are compared to RKE N$^4$LO after evolution to several values of $\lambda$ with $G = H_D$.
	The dotted lines indicate the value of $|\psi^A_{\infty}(q)|^2 / |\psi^B_{\infty}(q)|^2$.}
	\label{fig:R_ratio_comparison}
\end{figure*}

If we apply Eq.~\eqref{eq:factored_op} to matrix elements of $O_Q$ for the same Hamiltonian but in different nuclei, the integrations in the $\QQ$ block will be the same, so matrix-element ratios will be determined by soft (``mean-field'') physics and be independent of high-momentum details~\cite{Anderson:2010aq} (up to higher-order corrections beyond \eqref{eq:factored_op}). 
This explains why the high-momentum or short-distance behavior of momentum distributions is universal in nuclei~\cite{Anderson:2010aq,Bogner:2012zm,Neff:2015xda,Cruz-Torres:2019fum}.

\begin{figure*}[tbh]
	\includegraphics[clip,width=0.75\textwidth]{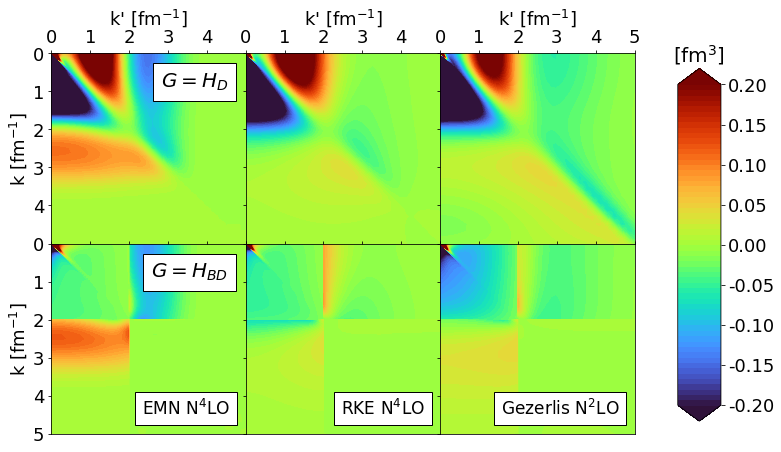}%
	\caption{Matrix elements of $\delta U(k,k')$ with the EMN N$^4$LO 500\,MeV, RKE N$^4$LO 450\,MeV, and Gezerlis \textit{et al.}~N$^2$LO 1\,fm potentials with Wegner and block-diagonal generators in the $^3$S$_1$ channel. Here we set $\lambda=1.5$\,fm$^{-1}$ for band-diagonal evolution, and $\LambdaBD=2$ and $\lambda=1$\,fm$^{-1}$ for block-diagonal evolution.}
	\label{fig:delta_U_default_potentials}
\end{figure*}

We can also use \eqref{eq:factored_op} to make comparisons for the same nucleus but with different potentials.
We use $\ataq$ with $q \gg \lambda$ or $\LambdaBD$ as an example, so 
\begin{align}
    \mel{q'}{[O_Q]_\infty}{q''} \rightarrow \delta(q'-q)\delta(q''-q)
\end{align}
and  the ratio for two different potentials A and B is
\begin{align}
	\label{eq:momentum_proj_matrix_elements_ratio}
	R[\ataq(k,k')] &\equiv
	\frac{\mel{k}{\bigl[\ataq\bigr]^A_{\lambda}}{k'}}
	{\mel{k}{\bigl[\ataq\bigr]^B_{\lambda}}{k'}} 
	\notag \\
   &\approx \frac{\Klo^A(k)\Klo^A(k') \Khi^A(q)^2}
             {\Klo^B(k)\Klo^B(k') \Khi^B(q)^2}
   .
\end{align}
From Fig.~\ref{fig:factorization_ratios} we verify that the $\Klo$ functions should be approximately the same for several chiral EFT potentials, so this ratio should be roughly constant for $k,k' < \lambda$ and $q \gg \lambda$.
This is illustrated in Fig.~\ref{fig:R_ratio_comparison} for these same potentials in the $^3$S$_1$ channel for $q = 3\,\mbox{fm}^{-1}$.
After evolution to $\lambda \ll q$, the ratio $R$ is quite flat in the unshaded region.
The value of the ratio at $q$ is well approximated at lower $\lambda$ by
\begin{align}
    \label{eq:momentum_proj_factorization_ratio}
    \frac{|\psi^A_{\infty}(q)|^2}{|\psi^B_{\infty}(q)|^2} =
	\frac{\mel{\psi^A_\lambda}{\bigl[\ataq\bigr]^A_{\lambda}}{\psi^A_\lambda}}
	{\mel{\psi^B_\lambda}{\bigl[\ataq\bigr]^B_{\lambda}}{\psi^B_\lambda}} 
	\approx \frac{\Khi^A(q)^2}
             { \Khi^B(q)^2}
    \equiv f(q),
\end{align}
where $\psi$ denotes the various deuteron wave functions, which share the same low-momentum structure so that the $\Klo$ dependence roughly cancels.

The high-momentum function $f(q)$ is dependent on the differences in the UV behavior of each of the representative potentials.
This does not mean, however, that one of the potentials is correct and the others are wrong.
Indeed, matrix elements of the representative operators considered in this section, $\ataq$ and $r^2$, cannot be absolutely measured by themselves in experiments. 
To relate them to measurable quantities, one must build and calibrate the initial operators for particular experimental observables, as done with EFTs.
A recent example is the precision calculation of the deuteron structure radius in Ref.~\cite{Filin:2019eoe}, which requires the inclusion of two-body currents that will have scale and scheme dependent contributions to match the measured charge form factor.
Only after the consistent construction of the Hamiltonian and current operators, with an assessment of uncertainties from theory discrepancies (such as EFT truncation errors), can one reliably compare predictions. 

\begin{figure*}[tbh]
	\includegraphics[clip,width=0.75\textwidth]{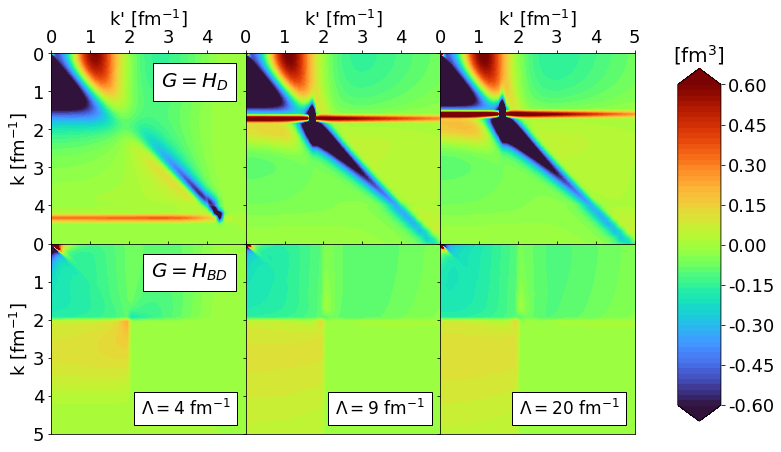}%
    \caption{Matrix elements of $\delta U(k,k')$ with the high-cutoff LO potentials with Wegner and block-diagonal generators in the $^3$S$_1$ channel. Here we set $\lambda=1.2$\,fm$^{-1}$ for band-diagonal evolution, and $\LambdaBD=2$ and $\lambda=1$\,fm$^{-1}$ for block-diagonal evolution.}
	\label{fig:delta_U_high_cutoffs}
\end{figure*}

For visual insight into factorization, we consider the SRG transformation directly.
We can write the SRG unitary transformation as
\begin{eqnarray}
    \label{eq:delta_U}
    U(s) = \mathbb{1} + \delta U(s),
\end{eqnarray}
where $\delta U(s)$ is responsible for the induced changes in transformed operators.
In Fig.~\ref{fig:delta_U_default_potentials} we show contours of $\delta U(s)$ in momentum space in the $^{3}S_1$ channel for the three representative potentials all evolved to $\lambda=1.5$\,fm$^{-1}$ with $G=H_D$ in the top row, and to $\lambda=1$ and $\LambdaBD=2$\,fm$^{-1}$ with $G=H_{BD}$ in the bottom row.
Figure~\ref{fig:delta_U_default_potentials} depicts factorization in the following sense.
By fixing $k'$ to a value much higher than $\lambda$ or $\LambdaBD$, we can take vertical lines up to $k=\lambda$ or $\LambdaBD$ and see little to no variation in the transformation.
The transformation approximately depends only on a function of high momentum $\Khi(k')$ in these regions, hence the same shade of color.
One can verify factorization in the opposite block by fixing $k \gg \lambda$ and taking horizontal lines up to $k'=\lambda$.

The difference in regulator functions between the three representative potentials is apparent in these figures.
For instance, in the EMN N$^4$LO case, the non-local regulator kills off the high-momentum matrix elements of $\delta U(s)$ because $U(s)$ is expressed in terms of the SRG generator $\eta(s)$ which contains the potential $V(k,k')$.
In the subsequent panels, the local momentum dependence of RKE N$^4$LO (semi-local) and Gezerlis N$^2$LO (local) is seen in non-zero matrix elements at higher momentum values.

Figure \ref{fig:delta_U_high_cutoffs} shows matrix elements of $\delta U(s)$ for the high-cutoff LO potentials.
These potentials still exhibit factorization of the SRG transformation, although the function of high momentum $\Khi(q)$ will have much different behavior than the softer chiral potentials.
Furthermore, we see the appearance of positive, horizontal bands for the band-diagonal transformations.
In the first column, the band corresponds to the region in which the high-momentum contributions of the initial potential accumulate up to the momentum space cutoff $\Lambda=4$\,fm$^{-1}$.
For $\Lambda=9$ and $20$\,fm$^{-1}$, the large and positive band corresponds to the decoupled spurious bound state.
This piece of the transformation decouples the spurious bound state along the diagonal in the evolved potential.
Otherwise, the low-momentum matrix elements are quite similar to the softer transformations in Fig.~\ref{fig:delta_U_default_potentials}, with larger contributions at high momentum due to the high momentum-space cutoffs.

In this section, we examined the characteristics of SRG evolution for representative operators $\ataq$ and $r^2$, extending the treatment in Ref.~\cite{Anderson:2010aq}.
The SRG changes in the operator do not lead to any complications that would offset the desired features in the decoupled NN potential.
Corresponding wave functions are decoupled in momentum space, either collapsing locally to one region with band-diagonal evolution, or cleanly cut off from low- or high-momentum sub-spaces with block-diagonal evolution (depending on the energy of the state).
The matrix elements of expectation values using the evolved operators and states show how one can take advantage of scale and scheme dependence to calculate consistent observable quantities at lower resolution.
We used factorization to show that high-momentum operators exhibit universal scaling dependent only on the high-momentum physics of the underlying NN potential.
That is, the high-momentum (short-distance) physics in the initial wave function appears in the evolved operator.

While the results of operator evolution generally have the same behavior for different potentials, the underlying scheme dependence is apparent.
The induced high-momentum (short-distance) contributions in non-Hamiltonian operators are dependent on the UV scheme of the NN potential.
Furthermore, the remaining SRG components of non-Hamiltonian operators reflect the scheme dependence of the bare operator.
For example, contrasting smeared delta functions to the single-point delta functions in $\ataq$ from \cite{Anderson:2010aq} gives differing behavior in the SRG-evolved matrix elements at $k,k'=q$.
Lastly, we demonstrated that the characteristics of operator evolution are the same for band- and block-diagonal SRG schemes, but the evolved matrix elements are different and reflect the SRG decoupling scheme.

\section{Summary and outlook}
\label{sec:summary}

Operator evolution is a critical aspect of SRG evolution in free-space and in-medium implementations.
Our initial focus here was on a technical aspect of this evolution: the efficacy and robustness of the Magnus expansion, which we evaluated in a difficult test environment with large cutoffs and spurious states.
But this study led us to reconsider and expand past studies of operator evolution in light of scheme dependencies arising both from different SRG generators and from different regulators of recent chiral EFT Hamiltonians.
The flow to low resolutions leads to universality (in the sense of independence from initial scheme dependence) in Hamiltonians and low-energy wave functions.
The constraint of unitarity then has implications for the corresponding flow of operators, while the scale separation from decoupling leads to consequences from factorization.
Each of these aspects can be exploited in future analyses of nuclear reactions that account for scale and scheme dependence.

We first used high-cutoff EFTs to verify that the Magnus expansion offers an improved variant of the standard SRG solution methods.
The Magnus implementation performs SRG transformations to exact unitarity which allows one to solve the flow equation \eqref{eq:srg_flow} using simple, efficient methods.
In Fock space, the benefits of the Magnus expansion are especially important as evolution of several operators simultaneously can be quite difficult for many-body systems, hence the prominence of the Magnus expansion in IMSRG calculations.
Here we showed that the Magnus implementation works effectively for a difficult free-space test problem in decoupling bound states using LO chiral potentials at high cutoffs.
The Magnus expansion reproduces the generator dependence seen in Ref.~\cite{Wendt:2011qj} and obtains eigen-energies to high accuracy.
In carrying out this test, we also showed that the block-diagonal generator decouples spurious, deeply bound states cleanly in the high-momentum sub-block but at higher momentum values than the band-diagonal generators and without apparent dependence on the discretization mesh.
However, the Magnus expansion does not converge in some cases, which is similar to a related issue in IMSRG calculations involving intruder states.
The NN convergence problem is related to the interaction and can be avoided in at least some cases by a careful selection of generator $G$.

The initial high-cutoff LO Hamiltonians and the contrasting evolution with band- and block-diagonal generators represent extremes of scale and scheme dependence.
Recently introduced chiral EFT Hamiltonians are characterized by a different type of scheme dependence in the use of qualitatively distinct regulators. 
In comparing their flows to low resolution, we confirmed that the momentum-space matrix elements of this new generation of \chiEFT\ Hamiltonians flow to a universal form when the decoupling scale is below the region of phase equivalence.
This happens for either band or block diagonal generators, but the universal form is not the same~\cite{Dainton:2013axa}.
We found small deviations from universality in channels dominated by one-pion exchange or only contact forces, which are attributed to the difference in regulator functions.
This was examined quantitatively using the Frobenius norm and SDT correlation coefficient and angle $\theta_{V,V'}$ as measures of the differences in the evolved potentials.

The flow to near-universality for potentials leads to almost perfect universality of deuteron wave functions.
Dramatically different initial wave functions in both their S-wave and D-wave characteristics collapse to near-indistinguishable low-resolution versions. 
We expect a similar collapse, if not as extreme, for the lowest-energy states in other nuclei.
This is encouraging for our goal of a controlled understanding of how spectroscopic factors are quenched in terms of a mismatch of high-resolution reaction models and low-resolution structure.
With universal wave functions at low resolution, we expect to identify universal features in the evolved reaction operators.

This goal led us to revisit the SRG evolution of non-Hamiltonian operators first studied in \cite{Anderson:2010aq,Bogner:2012zm} (see also \cite{Schuster:2013sda,Schuster:2014lga}).
We first extended our SRG analysis to momentum projection operators at low and high momentum.
Evolution of the momentum projection operator exemplifies the benefits that arise from SRG-transformed operators.
In particular, with decoupling at lower resolution there is a shift of strength to low momentum in matrix element of low-energy wave functions through induced two-body contributions (and smaller higher-body contributions that do not contribute to the deuteron).
This induced structure is very smooth and does not exhibit artifacts from the discretization of the operator.
The smoothness and universal properties are well understood from the factorization of the unitary transformations for well-separated momentum arguments.
The ratios of the same hard operators for different potentials scale with the high momentum and differ in magnitude as expected from differences in the ultraviolet content of the potentials.
(Note that to get the \emph{same} matrix elements, the operators themselves would have to be appropriately matched for the experimental observable in question.)
This suggests that a reliable theoretical understanding of high-energy reactions is possible using low-energy structure components (the initial wave function) with no insurmountable complications from the evolved operators.

Another representative operator is $r^2$, which is sensitive to the long-distance wave-function structure in coordinate space.
In momentum space, this operator has strength at all momentum scales and its visual form is highly sensitive to the momentum discretization scheme.
The two-body induced contributions from this operator to a low-energy state like the deuteron are small (see Ref.~\cite{Schuster:2014lga} for results on induced three-body contribution).
We isolated four types of contribution to the induced operator (see Fig.~\ref{fig:r2_srg_changes}).
The part originating fully from the high-momentum sector takes the same smooth form as the induced two-body operator for the high-$q$ momentum projection operator, and is explained in the same way by factorization.
However while the qualitative behavior of the other pieces is still explained by factorization, the numerical contributions are scheme dependent (see Table~\ref{tab:evolved_r2_contributions}).
This implies that even roughly estimating the net contribution, as in Ref.~\cite{Miller:2018mfb}, may be difficult.

The features highlighted here and in work on the electrodisintegration of the deuteron~\cite{More:2015tpa,More:2017syr} on the interplay of structure (wave functions) and reaction (operators) are promising for a cleaner theoretical understanding of FRIB-type knock-out reactions~\cite{Moro:2018mdk}.
By exploiting the unitary invariance of measured observables, we can shift the focus from correcting many-body wave functions to the computationally simpler RG flow of the operators. 
The long-standing and well-documented mismatch of experimental and theoretical cross sections for knock-out reactions (see Ref.~\cite{Tostevin:2014usa} and references therein) can be understood at least in part as a failure to do consistent matching of resolution scales.
That is, the over-prediction of cross sections with (low resolution) shell model wave functions should be understood as arising because a high-resolution reaction mechanism is used in the analysis. 
Exploiting the flow to universal soft wave functions and the corresponding consistent operators can open the door to process-independent analyses of these reactions.

\begin{acknowledgments}
We thank A. Garcia, P. Millican, and X. Zhang for fruitful discussions and feedback.
We are grateful to E. Epelbaum, R. Machleidt, and I. Tews for providing subroutines for calculating the chiral EFT NN potentials used in this work.
The work of AJT and RJF was supported by the National Science Foundation under Grant Nos.~PHY--1614460 and PHY--1913069, and the NUCLEI SciDAC Collaboration under US Department of Energy MSU subcontract RC107839-OSU\@.
The work of SKB was supported by the National Science Foundation under Grant No. PHY-1713901 and the U.S. Department of Energy, Office of Science, Office of Nuclear Physics under Grant No. de-sc0018083 (NUCLEI SciDAC Collaboration).
\end{acknowledgments}

\bibliography{tropiano_bib}

\end{document}